\documentclass[prodmode]{acmsmall} 

\usepackage[ruled]{algorithm2e}

\SetAlFnt{\small}
\SetAlCapFnt{\small}
\SetAlCapNameFnt{\small}
\SetAlCapHSkip{0pt}
\IncMargin{-\parindent}
\usepackage{subfigure} 

\usepackage{amssymb,amsbsy,amsmath,mathrsfs}
\usepackage[generate,shellescape,ps2eps]{abc}
\acmVolume{0}
\acmNumber{0}
\acmArticle{0}
\acmYear{0}
\acmMonth{0}

\newcommand{\vx}{\mathbf x}

\newcommand{\oneb}{\mathbf 1}

\newcommand{\vb}{\mathbf b}

\newcommand{\ve}{\mathbf e}

\newcommand{\vm}{\mathbf m}

\newcommand{\vy}{\mathbf y}

\newcommand{\MW}{\mathbf W}

\begin{document}

\markboth{Bob L. Sturm}{The ``Horse'' Inside}

\title{The ``Horse'' Inside: 
Seeking Causes Behind the Behaviours of Music Content Analysis Systems}
\author{Bob L. Sturm
\affil{School of Electronic Engineering and Computer Science, Queen Mary University of London}}

\begin{abstract}
Building systems that possess the sensitivity and intelligence to 
identify and describe high-level attributes in music audio signals 
continues to be an elusive goal,
but one that surely has broad and deep implications 
for a wide variety of applications.
Hundreds of papers have so far been published toward this goal,
and great progress appears to have been made.
Some systems produce remarkable accuracies 
at recognising high-level semantic concepts, such as music style, genre and mood.
However, it might be that these numbers do not mean what they seem.
In this paper, we take a state-of-the-art music content analysis system
and investigate what causes it to achieve
exceptionally high performance in a benchmark music audio dataset.
We dissect the system to understand its operation,
determine its sensitivities and limitations, and predict the kinds of knowledge
it could and could not possess about music.
We perform a series of experiments 
to illuminate what the system has actually learned to do,
and to what extent it is performing the intended music listening task.
Our results demonstrate how the initial manifestation of 
music intelligence in this state-of-the-art can be deceptive.
Our work provides constructive directions toward 
developing music content analysis systems that 
can address the music information and creation needs of real-world users.
\end{abstract}




\acmformat{Bob L. Sturm. 2016. The ``Horse'' Inside.}

\begin{bottomstuff}
Author's address: School of Electronic Engineering and Computer Science,
Queen Mary University of London,
Mile End Road London E1 4NS, UK.
\end{bottomstuff}

\maketitle

\section{Introduction}
A significant amount of research
in the disciplines of music content analysis
and content-based music information retrieval (MIR)
is plagued by an inability to distinguish between solutions and ``horses'' 
\cite{Gouyon2013,Urbano2013,Sturm2013g,Sturm2013h}.
In its most basic form, a ``horse'' is a system that {\em appears} as if it is
solving a particular problem when it actually is not \cite{Sturm2013g}.
This was exactly the case with Clever Hans \cite{Pfungst1911}, 
a {\em real} horse that was claimed to be capable of doing arithmetic
and other feats of abstract thought.
Clever Hans appeared to answer complex questions posed to him, 
but he had actually learned to respond to involuntary cues 
of his many inquisitors confounded with the tapping of his hoof
the correct number of times.
The ``trick'' evaded discovery for a few reasons:
1) the cues were nearly undetectable;
and 2) in light of undetected cues, the demonstration
was thought by many to constitute valid evidence 
for the claim that the horse possessed such abilities.
It was not until controlled experiments were designed and implemented 
that his true abilities were discovered \cite{Pfungst1911}.

If the aim of a music content analysis system is to 
enhance the connection between users 
(e.g., private listener, professional musician, scholar,
journalist, family, organisation and business) 
and music (e.g., recordings in the format of a score and audio recording)
and information about music (e.g., artist, tempi, instrumentation and title) \cite{Casey2008a}
-- and to do so at a far lower cost than that required of human labor -- 
then the system must 
operate with characteristics and criteria {\em relevant} to the information needs of users.
For instance, a relevant characteristic for generating tempo information
is periodic onsets; an irrelevant characteristic is instrumentation.
If the aim of a music content analysis system is to facilitate creative pursuits, 
such as composing or performing music in particular styles \cite{Dubnov2003a,Dubnov2014a},
then it must operate with characteristics and criteria {\em relevant} to 
the creative needs of users.
For instance, a relevant criterion for a Picardy third 
is the suggestion of a minor resolution;
an irrelevant criterion is avoidance of parallel fifths.
The importance of ``relevant criteria'' in music content analysis is evinced by
frustration surrounding what has been termed the ``semantic gap'':
a chasm of disconnection between accessible but low-level features 
and high-level abstract ideas \cite{Aucouturier2009,Wiggins2009,Turnbull2008}.

A music content analysis system's reproduction 
of dataset ground truth is, by and large, considered valid evidence
that the system is using relevant characteristics and criteria,
or possesses ``musical knowledge,''
or has learned to listen to music
in a way that is meaningful with respect to some
music listening task. 
In one of the most cited papers in MIR, \citeN{Tzanetakis2002}
train and test several systems with what would become the most-used
public benchmark dataset in music genre recognition \cite{Sturm2013h}.
Since these systems reproduced an amount of ground truth
inconsistent with that expected when choosing classes randomly, 
Tzanetakis and Cook concluded that the features ``provide some information about 
musical genre and therefore musical content in general,'' 
and even that the systems' performances are comparable to that of humans \citeyear{Tzanetakis2002}.
 \citeN{Tsunoo2009b} conclude from such evidence that
 the features they propose
``have enough information for genre classification 
because [classification accuracy] is significantly above the baselines of random classification.''
Measuring the reproduction of the ground truth of a dataset
is typical when developing content analysis systems.
For instance, \citeN{Song2012} and \citeN{Su2014}
perform a large number of computational experiments
to find the ``most relevant'' features, ``optimal'' parameters, ``best'' classifiers, 
and combinations thereof, all defined with respect to 
the reproduction of the ground truth.

The measurement of reproduced ground truth 
has been thought to be objective. 
\citeN{Gouyon2004} avoid the ``pitfall'' of subjective evaluation
of rhythm descriptors by ``measuring their rate of success in genre classification experiments."
\citeN{Lidy2008} argue that such ``directly measured'' numbers ``[facilitate]
(1) the comparison of feature sets and 
(2) the assessment of the suitability of particular classifiers for specific feature sets.''
This is also echoed by \citeN{Tzanetakis2002}.\footnote{Also see lecture by G. Tzanetakis, ``UVic MIR Course'': 
\url{https://www.youtube.com/watch?v=vD5wn-ffVQY} (2014).}
During the 10-year life-span of MIREX --
an established annual event 
that facilitates the exchange and scientific evaluation of new techniques 
for a wide variety of music content analysis tasks 
\cite{Downie2004b,Downie2008,Downie2010,Cunningham2012} --
thousands of systems have been 
ranked according to the amount of ground truth they reproduce.
Several literature reviews, e.g., \cite{Scaringella2006,Fu2011,Humphrey2013},
tabulate results of many published experiments,
and make conclusions about which features and classifiers are ``useful''
for listening tasks such as music genre and mood classification.
\citeN{Bergstra2006} remark on the progress up to that time, 
``Given the steady and significant improvement in classification [accuracy], 
we wonder if automatic methods are not already 
more efficient at learning genres than some people.''
Seven years later, \citeN{Humphrey2013} surmise from the plateauing of
such numbers that progress in MIR has stalled.
However, could it be that progress was never made at all? 
Might it be that the precise measurement of reproduced ground truth 
is not a reliable reflection of the ``intelligence'' so hoped for?

\begin{figure}[t]
\centering
\includegraphics[width=2.8in]{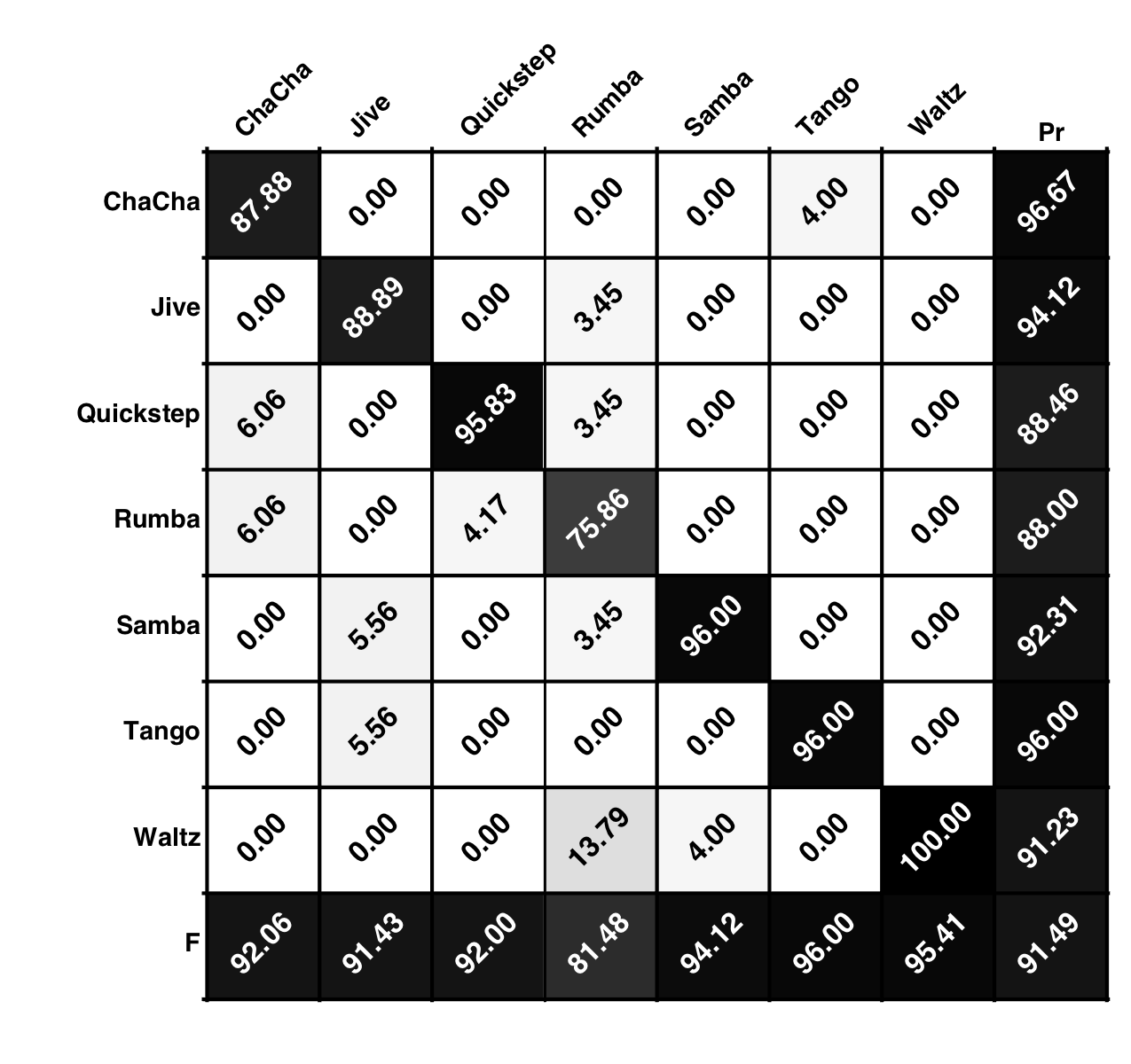}
\caption{Figure of merit (FoM, $\times 100$) of the 
music content analysis system DeSPerF-BALLROOM,
the cause of which we seek in this article.
Column is ground truth label, and row is class selected by system.
Off diagonals are confusions.
Precision is the right-most column,
F-score is the bottom row,
recall is the diagonal,
and normalised accuracy (mean recall) is at bottom-right corner.}
\label{fig:DeSPerF_expt00}
\end{figure}

Consider the systems reproducing the most ground truth in the
2013 MIREX edition of the ``Audio Latin Music Genre classification task'' (ALGC).\footnote{\url{http://www.music-ir.org/nema_out/mirex2013/results/act/latin_report/}}
The aim of ALGC is to compare music content analysis systems 
built from algorithms submitted by participants
in the task of classifying the music genres 
of recordings in the benchmark Latin Music Dataset ({\em LMD}) \cite{Silla2008b}.
In ALGC, participants submit their feature extraction and machine learning algorithms,
a MIREX organiser then uses these to build music content analysis systems,
applies them to subsets of {\em LMD}, 
and computes a variety of figures of merit (FoM) based on the reproduction of ground truth.
In ALGC of 2013, the most ground truth (accuracy of $0.776$) was reproduced
by systems built using deep learning \cite{Pikrakis2013}.
Figure \ref{fig:DeSPerF_expt00} shows the FoM of the system
resulting from using the same winning algorithms, but training and testing it with 
the public benchmark {\em BALLROOM} dataset \cite{Dixon2004}.
({\em LMD} is not public.)

Of little doubt is that the classification accuracy of this system -- DeSPerF-BALLROOM --
greatly exceeds that expected when selecting labels of
{\em BALLROOM} randomly.
The system has clearly learned something.
Now, {\em what is that something}? 
What musical characteristics and criteria -- ``musical knowledge'' -- is this system using?
How do the internal models of the system reflect the music ``styles'' in {\em BALLROOM}?
Are the labels of {\em BALLROOM} even related to ``style''?
What has this system {\em actually} learned to do with {\em BALLROOM}?
The success of DeSPerF-BALLROOM
for the analytic or creative objectives of music content analysis
turns on the {\em cause} of Fig. \ref{fig:DeSPerF_expt00}.
How is the cause relevant to a user's music information or creation needs?
Is the system {\em actually} fit to
enhance the connections between users, music,
and information about music?
Or is it as Clever Hans,
only appearing to be intelligent?

In this article, it is the 
cause of Fig. \ref{fig:DeSPerF_expt00}
with which we are principally concerned.
We seek to answer what this 
system has learned about the music it is classifying, its {\em musical intelligence},
i.e., its decision machinery involving high-level 
acoustic and musical characteristics of the ``styles''
from which the recordings {\em BALLROOM} appear to be sampled.
Broader still, we seek to encourage completely new methods for 
evaluating any music content analysis system
with respect to its objective.
It would have been a simple matter if Hans could have been asked  
how he was accomplishing his feat; but the nature of his ``condition''
allowed only certain questions to be asked.
In the end, it was not about finding the definitive set of questions
that accurately measured his mental brawn, 
but of thinking skeptically and implementing appropriately controlled experiments
designed to test hypotheses like, ``Clever Hans can solve problems of arithmetic.''
One faces the same problem in evaluating music content analysis systems:
the kinds of questions that can be asked are limited.
For DeSPerF-BALLROOM in Fig. \ref{fig:DeSPerF_expt00},
a ``question'' must come in the form of a 220,500-dimensional vector
(10 second monophonic acoustic music signal
uniformly sampled at 22050 Hz).
Having the system try to classify the 
Waltz recordings that are thought to be the hardest in some way
will not illuminate much about the criteria it is using,
the sanity of its internal models, or the causes of the FoM in Fig. \ref{fig:DeSPerF_expt00}.
Ways forward are given by adopting and adapting 
Pfungst's approach to testing Clever Hans \cite{Pfungst1911},
and above all not dispensing with skepticism.

Teaching a machine to listen to music,
to automatically recognise music style or genre,
are achievements so great that they require extraordinary and valid evidence.
That these tasks defy the explicit definition
necessary to the formal nature of algorithms
produces great pause in accepting Fig. \ref{fig:DeSPerF_expt00}
as evidence that DeSPerF-BALLROOM is, 
unlike Clever Hans, not a ``horse.''
In Section \ref{sec:problemofMCA}, we provide a brief but explicit definition 
of the problem of music content analysis,
and what a music content analysis system is.
In Section \ref{sec:DeSPerF}, we dissect DeSPerF-based systems,
describing in detail their construction and operation.
In Section \ref{sec:BALLROOM}, we analyse the methods of teaching and testing
encompassed by the {\em BALLROOM} dataset.
We are then prepared for the series of experiments in Section \ref{sec:experiments}
that seek to explain Fig. \ref{fig:DeSPerF_expt00}.
We discuss our results more broadly in Section \ref{sec:discussion}.
We make available a reproducible research package
with which one may generate all figures and tables
in this article: \url{http://manentail.com}.
(Made anonymous for the time being.)

\newcommand{\UniverseMusic}{{\Omega}} 
\newcommand{\UniverseRecording}{{\mathcal{R}_\UniverseMusic}} 
\newcommand{\Vocabulary}{{\mathcal V}} 
\newcommand{\FeatureVocabulary}{{\mathbb F }} 
\newcommand{\UniverseSemantic}{{\mathcal U}_{\Vocabulary,A}} 
\newcommand{\UniverseSemanticFeature}{{\mathcal{U}_{\mathbb{F},A'}}} 
\newcommand{\recording}{{r_\omega}} 
\newcommand{\system}{{\mathscr{S}}} 

\section{The problem of music content analysis}\label{sec:problemofMCA}
Since this article is concerned with 
algorithms defined in no uncertain terms by a formal language
and posed to solve some problem of music content analysis,
we must define what all these things are.
Denote the {\em music universe} $\UniverseMusic$,
the {\em music recording universe} $\UniverseRecording$
(notated or performed),
{\em vocabularies} $\FeatureVocabulary$ (features) and $\Vocabulary$ (tokens),
and define the {\em Boolean semantic rules} $A': f \to \{T,F\}$ and $A: s \to \{T,F\}$,
where $f$ is a sequence of features from $\FeatureVocabulary$ 
and $s$ is a sequence of tokens from $\Vocabulary$.
Define the {\em semantic universe} built from $\Vocabulary$ and $A$:
\begin{equation}
\UniverseSemantic := \{s \in \mathcal{V}^n | 
n \in \mathbb{N} \land A(s) = T\}.
\end{equation}
The {\em semantic feature universe} $\UniverseSemanticFeature$ is similarly built 
using $\FeatureVocabulary$ and $A'$. 
Define a {\em use case} as the specification of 
$\UniverseMusic$, $\UniverseRecording$, $\UniverseSemantic$,
and a set of success criteria.

A music universe $\UniverseMusic$ is the set of intangible music -- whatever that is \cite{Goehr1994} --
from which the tangible recording music universe $\UniverseRecording$ is produced.
This distinction is important because the real world contains
only tangible records of music.
One can point to a score of Beethoven's 5th,
but not to Beethoven's 5th.
Perhaps one wishes to say something about
Beethoven's 5th, or about a recording of Beethoven's 5th.
These are categorically different.
The definition of $\UniverseMusic$ specifies
the music in a use case, e.g., ``music people call `disco'.''
The definition of $\UniverseRecording$ includes the specification of
the dimensions of the tangible material,
``30 second audio recording uniformly sampled at 44.1 kHz
of an element of $\UniverseMusic$.''
The definition of $\UniverseSemantic$ provides the 
semantic space in which elements of $\UniverseRecording$ 
are described.
Finally, the success criteria of a use case specify requirements
for music content analysis systems to be deemed successful.

A {\em music content analysis system} $\system$ is a map
from $\UniverseRecording$ to $\UniverseSemantic$:
\begin{equation}
\system : \UniverseRecording \rightarrow \UniverseSemantic
\end{equation}
which itself is a composition of two maps,
$\mathscr{E}: \UniverseRecording \rightarrow \UniverseSemanticFeature$
and $\mathscr{C}: \UniverseSemanticFeature \rightarrow \UniverseSemantic$.
The map $\mathscr{E}$ is commonly known as a ``feature extractor,''
taking $\UniverseRecording$
to $\UniverseSemanticFeature$;
the map $\mathscr{C}$ is commonly known as a ``classifier'' or ``regression function,''
mapping $\UniverseSemanticFeature$ to $\UniverseSemantic$.
The {\em problem of music content analysis} is to 
build a system $\system$ that meets the success criteria of a use case.
A typical procedure for building an $\mathscr{S}$ is to seek a way to reproduce
all the ground truth of a {\em recorded music dataset}, 
defined as an indexed sequence 
of tuples sampled in some way from the population 
$\UniverseRecording\times\UniverseSemantic$, i.e.,
\begin{equation}
\mathcal{D} := \left( (r_i,s_i) : i \in \mathcal{I} \right ) \subset R_\Omega\times \UniverseSemantic
\end{equation}
where $\mathcal{I}$ indexes the dataset.
We call $(s_i)_{i \in \mathcal{I}}$ the {\em ground truth} of $\mathcal{D}$.

As a concrete example, take the {\em Shazam} 
music content analysis system \cite{Wang2003}.\footnote{\url{http://www.shazam.com/}}
One can define its use case as follows.
$\UniverseRecording$ and $\UniverseMusic$ are defined entirely from
the digitised music recordings in the Shazam database.
$\UniverseMusic$ is defined as the set of music {\em exactly} as it appears 
in specific recordings.
$\UniverseRecording$ is defined by all 10-second audio recordings
of elements of $\UniverseMusic$.
$\UniverseSemantic$ is defined as a set of single tokens,
each token consisting of an artist name, song title, album title, and other metadata.
The {\em Shazam} music content analysis system maps a 10 second audio recording of $\UniverseRecording$
to an element of $\UniverseSemanticFeature$ consisting of many 
tuples of time-frequency anchors $\UniverseSemanticFeature$.
The classifier then finds matching time-frequency anchors
in a database of all time-frequency anchors from $\UniverseRecording$,
and finally picks an element of $\UniverseSemantic$.
The success criteria might include making correct mappings
(retrieving the correct song and artist name of the specific music heard)
in adverse recording conditions, or increased revenue from music sales.

\section{DeSPerF-based Music Content Analysis Systems}\label{sec:DeSPerF}
In the following subsections, we dissect DeSPerF-BALLROOM,
first analysing its feature extraction, and then its classifier.
This helps determine its sensitivities and limitations.
The feature extraction of DeSPerF-based systems 
maps $\UniverseRecording$ to $\UniverseSemanticFeature$,
using {\em spectral periodicity features} (SPerF),
first proposed by \citeN{Pikrakis2013}.
Its classifier maps $\UniverseSemanticFeature$ to $\UniverseSemantic$
using 
deep neural networks (DNN).
In the case of DeSPerF-BALLROOM in Fig. \ref{fig:DeSPerF_expt00}, 
$\UniverseSemantic := \{$``Cha cha'', ``Jive'', ``Quickstep'', ``Rumba'', ``Tango'', ``Waltz''$\}$.

\newcommand{\durSegment}{{T_\textrm{seg}}} 
\newcommand{\durSegmentSkip}{{T_\textrm{seghop}}} 
\newcommand{\durFrame}{{T_\textrm{fr}}} 
\newcommand{\durFrameSkip}{{T_\textrm{frhop}}}
\newcommand{\numMFCCs}{{N_\textrm{MFCCs}}}
\newcommand{\numframes}{{N_\textrm{fr}}}

\subsection{Feature extraction}\label{sec:features}
SPerF describe temporal periodicities of modulation sonograms.
The hope is that SPerF reflect, or are correlated with, high-level
musical characteristics such as tempo, meter and rhythm \cite{Pikrakis2013}.
The feature extraction is defined by
six parameters: $\{\durSegment, \durSegmentSkip, \durFrame, \durFrameSkip, \numMFCCs, \numframes\}$.
It takes an element of $\UniverseRecording$
and partitions it into multiple {\em signal segments} of duration $\durSegment$ seconds (s) 
which {\em hop} by $\durSegmentSkip$ s. 
Each signal segment is divided into {\em frames} of duration $\durFrame$ s 
with a hop of $\durFrameSkip$ s.
From the ordered frames of a segment, 
a sequence of the first $\numMFCCs$ Mel-frequency cepstral coefficients (MFCCs) are computed, 
which we call the {\em segment modulation sonogram} 
\begin{equation}
\mathcal{M} = \left (\vm_i \in \mathbb{R}^{\numMFCCs} : 
i \in [0, \ldots, {\max}_i] \right )
\label{eq:segmodsonogram}
\end{equation}
where $\vm_i $ is a vector of MFCCs
extracted from the frame spanning time $[i\durFrameSkip,i\durFrameSkip+\durFrame]$,
and ${\max}_i := \lfloor (\durSegment-\durFrame)/\durFrameSkip\rfloor +1$
is the index of the last vector.

\begin{figure}[t]
\centering
\includegraphics[width=4in]{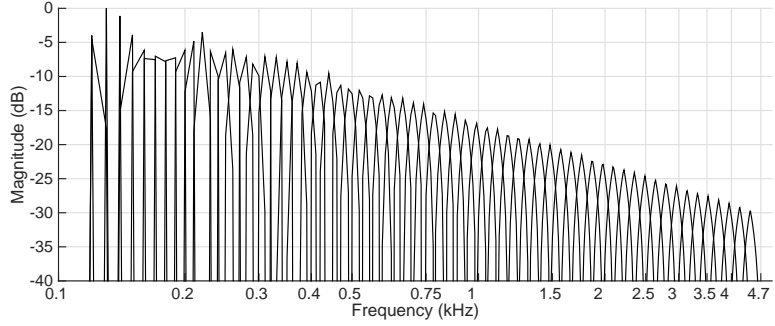}
\caption{MFCC filterbank used by the feature extraction of 
DeSPerF-based systems.}
\label{fig:SPerFMFCC}
\end{figure}

The MFCCs of a frame are computed by a modification of the approach of \citeN{Slaney1998}.
The magnitude discrete Fourier transform (DFT) of a Hamming-windowed frame 
is weighted by a ``filterbank'' of 64 triangular filters,
the centre frequencies of which are spaced by one semitone.
Figure \ref{fig:SPerFMFCC} shows these filters.
Each filter is weighted inversely proportional to its bandwidth.
The lowest centre frequency is 110 Hz, and the highest is 4.43 kHz.
Irregularities in filter shape at low frequencies
arise from the uniform resolution of the DFT 
and the frame duration $\durFrame$ s.
Finally, the discrete cosine transform (DCT) 
of the $\log_{10}$ rectified filterbank output is taken,
and the first $\numMFCCs$ MFCCs are selected to form $\vm_i$.
The period corresponding to the $k$th MFCC is
$128/k$ semitones, $k \in \{1, \ldots, \numMFCCs-1\}$,
and $0$ for $k=0$.
The first MFCC ($k=0$) is related to the mean energy over all 64 semitones.
The third MFCC is related to the amount of 
energy of a component with a period of the entire filterbank.
And the eleventh MFCC is related to the amount of energy 
of a component with a period of an octave.

For each {\em lag} $l \in \{1, \ldots, \numframes\}$, 
define the two {\em lagged} modulation sonograms 
\begin{align}
\mathcal{M}_{:l} & = (\vm_i \in \mathcal{M} : i \in [0, {\max}_i-l]) \\
\mathcal{M}_{l:} & = (\vm_i \in \mathcal{M} :  i \in [l, {\max}_i]).
\end{align}
$\mathcal{M}_{:l}$ starts from the beginning of the segment,
and $\mathcal{M}_{l:}$ ends at the segment's conclusion.
A lag $l$ corresponds to a time-shift of $l\durFrameSkip$ s
between the sonograms.
Now, define the {\em mean distance} between these modulation sonograms at lag $l$ 
\begin{equation}
d[l] = \frac{\|\textrm{vec}(\mathcal{M}_{:l}) - \textrm{vec}(\mathcal{M}_{l:})\|_2}{|\mathcal{M}_{:l}|}
\label{eq:meandistancelaggedmodsonograms}
\end{equation}
where $\|\cdot \|_2$ is the Euclidean norm,
and $\textrm{vec}(\mathcal{M})$ stacks the ordered
elements of sequence $\mathcal{M}$ into a column vector.
The sequence $d[l]$ is then filtered,
$y[l] = \left ( (d * h) * h \right )[l]$,
where
\begin{equation}
h[n] = \begin{cases}
\frac{1}{n}, & -\durFrame/\durFrameSkip \le n \in \mathbb{Z}\backslash 0 \le \durFrame/\durFrameSkip \\
0, & \textrm{otherwise}
\end{cases}
\end{equation}
and adapting $h[n]$ around the end points of $d[l]$
(shortening its support to a minimum of two).
This sequence $y[l]$ approximates the second derivate of $d[l]$.
Finally, a SPerF of an audio segment 
is created by the sigmoid normalisation of $y[l]$:
\begin{equation}
x[l] = [1 + \exp\left (- (y[l] - \hat \mu_y)/ \hat \sigma_y \right )]^{-1}, 1 \le l \le \numframes
\label{eq:SPerF}
\end{equation}
where $\hat \mu_y$ is the mean of $(y[l] : 1 \le l \le \numframes)$
and $\hat \sigma_y$ is its standard deviation.

The output of the feature extraction is a sequence $f$ of SPerFs (\ref{eq:SPerF}),
each element of which is computed from one segment of the recording $r$.
In this case, the feature vocabulary is defined
$\mathbb{F} := (0,1)^{\numframes}$.
The semantic rule is defined $A'(f) := (|f| \le (|r|_s-\durSegment)/\durSegmentSkip+1)$,
where $|r|_s$ is the duration of the recording from $\UniverseRecording$ in seconds.
Together, these define $\UniverseSemanticFeature$.
Table \ref{tab:featureExtraction} summarises the six parameters of the feature extraction and
their interpretation, as well as the values used in the system of Fig. \ref{fig:DeSPerF_expt00}.

\begin{table}%
\tbl{Parameters of the feature extraction algorithm $\mathscr{E}$ \label{tab:featureExtraction}
of the DeSPerF-based system of Fig. \ref{fig:DeSPerF_expt00}}{%
\begin{tabular}{|r|p{0.9in}|p{3.3in}|}
Symbol & Value & Interpretation \\
\hline
$\durSegment$ & 10s & Segment duration for computing SPerF from an element of $\UniverseRecording$. 
Limits time over which repetitive musical events can be detected. \\\
$\durSegmentSkip$ & 1s & Hop of each segment along an element of $\UniverseRecording$.
The elements of $\UniverseSemanticFeature$
become more redundant as $\durSegmentSkip \to 0$. \\ \hline
$\durFrame$ & 100 ms & Duration of frame in a segment in which MFCCs are computed. \\
$\durFrameSkip$ & 5 ms & Hop of each frame along a segment.
The dimensionality of $\mathbb{F}$
becomes larger as $\durFrameSkip \to 0$. \\\hline
$\numMFCCs$ & 13 & Number of MFCCs to compute for a frame, implemented in \cite{Slaney1998}.
Limits resolution of spectral structures considered in SPerF computation. \\
$\numframes$ & 800 & Maximum lag to consider. $l\durFrameSkip$ is the time lag, $l\in\{1,\ldots,\numframes\}$.
Note, $\numframes < {\max}_i := (\durSegment-\durFrame)/\durFrameSkip +1$.
Limits time over which repeated musical events can be detected.\\ \hline
\end{tabular}}
\begin{tabnote}%
\end{tabnote}%
\end{table}%

We can relate characteristics of a SPerF 
to low-level characteristics of a segment of a recording.
For instance, if $\mathcal{M}$ is periodic with
period $T$, then for $l \approx k T/\durFrameSkip, k \in \{1, 2, \ldots\}$
the mean distance sequence $d[l]$ should be small,
$y[l]$ should be large positive ($d[l]$ is convex around these lags), 
and $x[l]$ should be close to one.
If $\mathcal{M}$ is such that $d[l]$ is approximately constant over its support,
then $y[l]$ will be approximately zero, and $x[l] \approx 0.5$.
This is the case if a recording is silent or is not periodic
within the maximum lag $\numframes\durFrameSkip$ s.
If $x[l]$ is approximately zero at a lag $l$, 
then $y[l]$ is very negative,
and there is a large distance $d[l]$ between
lagged modulation spectrograms around that lag.

Moving to higher-level characteristics,
we can see that if the recording has a repeating timbral structure 
within the segment duration $\durSegment$ s,
and if these repetitions occur within $\numframes\durFrameSkip$ s,
then $x[l]$ should have peaks around those lags corresponding to the 
periodicity of those repetitions.
The mean difference between lags of successive peaks
might then be related to the mean tempo of music in the segment,
or at least the periodic repetition of some timbral structure.
If periodicities at longer time-scales exist in $x[l]$,
then these might be relatable to the meter of the music in the segment,
or at least a longer time-scale repetition of some timbral structure.

\begin{figure}[t]
\centering
\includegraphics[width=4.2in]{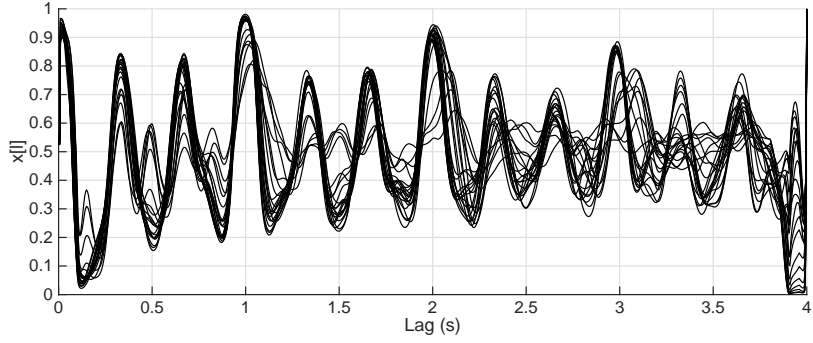}
\caption{Examples of SPerF (\ref{eq:SPerF}) extracted from 
{\em BALLROOM} Waltz recording {\em Albums-Chrisanne1-08}.}
\label{fig:SPerFexample}
\end{figure}

Figure \ref{fig:SPerFexample} shows several SPerF
extracted from recording 
of {\em BALLROOM}.
The SPerF shows a short-term periodicity of about $0.33$ s,
and about $1$ s between each of the first three highest peaks.
The tempo of the music in this recording is 
about 180 beats per minute (BPM), and it has a triple meter.
The few SPerF that do not follow the main trend
are from the introduction of the recording,
during which there are not many strong and regular onsets.

\subsection{Classification}
From a recording $r$,
$\mathscr{E}$ extracts a sequence of $N(r)$ SPerF, 
$f = (\vx_1, \vx_2, \ldots, \vx_{N(r)} )$,
where $\vx_j \in \mathbb{F}$ is a vectorised SPerF (\ref{eq:SPerF}).
The classifier $\mathscr{C}$ maps $f$ to
$\UniverseSemantic$ by a cascade of $K$ steps.
At step $1 \le k < K$, the $j$th SPerF $\vx_j^{(0)} \leftarrow \vx_j$ has been transformed iteratively by
\begin{equation}
\vx_j^{(k)} \leftarrow  \sigma \left(\MW_k\vx_j^{(k-1)} + \vb_k\right )
\label{eq:neuronoutput}
\end{equation}
where $\MW_k$ is a real matrix, $\vb_k$ is a real vector, and
\begin{equation}
\sigma(\vy) := \frac{1}{1+\exp(-\vy)}.
\label{eq:signmoid}
\end{equation}
Step $K$ produces a vector  
of posterior probabilities over $\UniverseSemantic$
by a softmax output
 \begin{equation}
\vx_j^{(K)} \leftarrow \frac{\exp \left [ \MW_{K}\vx_j^{(K-1)} + \vb_K \right ]}{\oneb^T\exp \left [ \MW_{K}\vx_j^{(K-1)} + \vb_K \right ]}
\label{eq:DNNoutput}
\end{equation}
where $\oneb$ is an appropriately sized vector of all ones.

The cascade from $\vx_j^{(0)}$ to $\vx_j^{(K)}$
is also known as a {\em deep neural network} (DNN),
with (\ref{eq:DNNoutput}) being interpreted as posterior probabilities 
over the sample space defined by $\UniverseSemantic$.
If all elements of $\vx_j^{(K)}$ are the same, 
then the DNN has no ``confidence'' in any particular element of $\UniverseSemantic$
given the observation $\vx_j$.
If all but one element of $\vx_j^{(K)}$ are zero,
then the DNN has the most confidence that
$\vx_j$ points only to a specific element of $\UniverseSemantic$.
Finally, $\mathscr{C}$ maps the sequence of posterior probabilities $(\vx_1^{(K)}, \ldots, \vx_{N(r)}^{(K)})$ 
to $\UniverseSemantic$ by {\em majority vote}, i.e.,
\begin{equation}
\hat s (f) := \arg \max_{s \in \UniverseSemantic} \sum_{j=1}^{N(r)}  I_{\UniverseSemantic}\left(\vx_j^{(K)}, s  \right) 
\end{equation}
where $I_{\UniverseSemantic}(\vx, s) = 1$ if $s$ 
is the element of $\UniverseSemantic$ associated with the largest value in $\vx$,
and zero otherwise.


The classifier of the system of Fig. \ref{fig:DeSPerF_expt00} has $K=6$ layers,
with the matrices and biases being:
$\MW_1 \in \mathbb{R}^{500\times800}$;
$\MW_2,\MW_3, \MW_4 \in \mathbb{R}^{500\times500}$;
$\MW_5 \in \mathbb{R}^{2000\times500}$;
$\MW_6 \in \mathbb{R}^{7\times 2000}$;
$\vb_1, \vb_2, \vb_3, \vb_4 \in \mathbb{R}^{500}$;
$\vb_5 \in \mathbb{R}^{2000}$;
and finally $\vb_6 \in \mathbb{R}^{7}$.
The set of parameters $\{\{\MW,\vb\}_k : 1\le k\le 6\}$ 
are found by using a training dataset and {\em deep learning} \cite{Deng2014}.\footnote{
For the system of Fig. \ref{fig:DeSPerF_expt00}
this is done by adapting the code 
produced by Salakhutdinov and Hinton 
(\url{http://www.cs.toronto.edu/~hinton/MatlabForSciencePaper.html}),
which trains a deep autoencoder for handwritten digit recognition.
This code for DeSPerF is provided by A. Pikrakis.}
Interpreting these parameters is not straightforward, 
save those at the input to 
the first hidden layer, i.e., $\MW_1$ and $\vb_1$.
The weights $\MW_1$ describe what information of a SPerF $\vx_j$
is passed to the hidden layers of the DNN.
The $m$th element of the vector $\MW_1\vx_j$ is the input to 
the $m$th neuron in the first hidden layer.
Hence, this neuron receives the product of the $m$th row of $\MW_1$ with $\vx_j$.
When those vectors point in nearly the same direction, 
this value will be positive;
when they point in nearly opposite directions,
this product will be negative;
and when they are nearly orthogonal, the product will be close to zero.
We might then interpret each row of $\MW_1$ as being
exemplary of some structures in SPerF
that the DNN has determined to be important for $\UniverseSemantic$.

\begin{figure}[t]
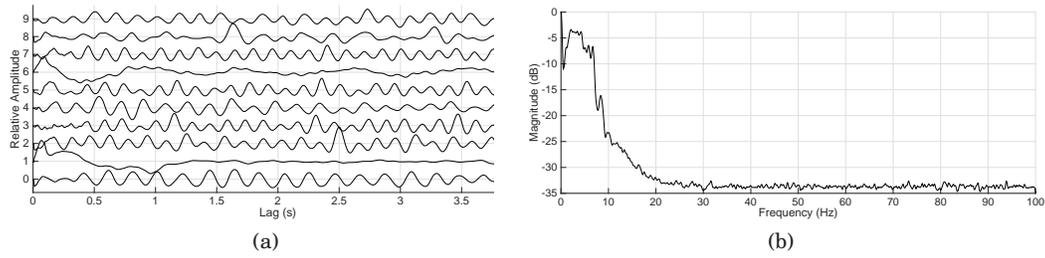

\centering
\subfigure[]{
\includegraphics[width=2.7in]{DeSPerF_BALLROOM_random_DNN_layer1.eps}
}\hspace{-0.1in}
\subfigure[]{
\includegraphics[width=2.7in]{DeSPerF_BALLROOM_random_DNN_layer1magnitude.eps}}
\caption{(a) The rows of $\MW_1$ with the largest Euclidean norm
from the system of Fig. \ref{fig:DeSPerF_expt00}.
(b) Combined magnitude response of all Hann windowed weights 
of the first layer of the DNN of the same system.}
\label{fig:DNNweights}
\end{figure}

Figure \ref{fig:DNNweights}(a) shows for DeSPerF-BALLROOM
the ten rows of $\MW_1$ with the largest Euclidean norm.
Many of them bear resemblance to the kinds of structures 
seen in the SPerF in Fig. \ref{fig:SPerFexample}.
We can determine the bandwidth of the input to the first hidden layer
by looking at the Hann-windowed rows of $\MW_1$ in the frequency domain.
Figure \ref{fig:DNNweights}(b) shows the sum of the 
magnitude responses of each row of $\MW_1$
for the system of Fig. \ref{fig:DeSPerF_expt00}.
We see that 
the majority of energy of a SPerF transmitted into the hidden layers of its DNN
is concentrated at frequencies below 10 Hz.

\begin{figure}[t]
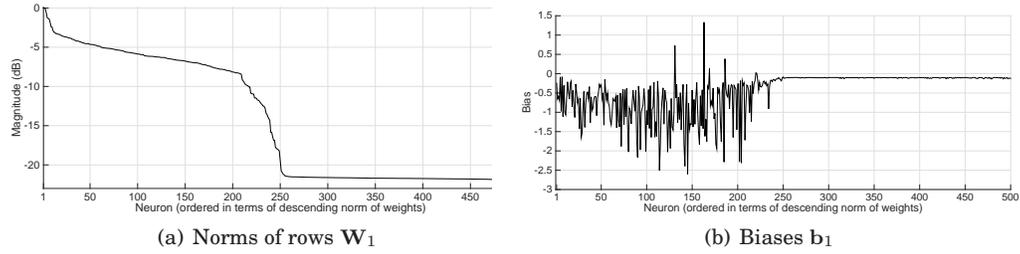

\centering
\subfigure[Norms of rows $\MW_1$]{
\includegraphics[width=2.7in]{DeSPerF_BALLROOM_random_DNN_layer1_norm.eps}}\hspace{-0.1in}
\subfigure[Biases $\vb_1$]{
\includegraphics[width=2.6in]{DeSPerF_BALLROOM_random_DNN_layer1_bias.eps}}
\caption{Characteristics of the parameters of the first layer $\MW_1, \vb_1$ 
of the DNN in the system of Fig. \ref{fig:DeSPerF_expt00}.}
\label{fig:DNNparameters}
\end{figure}

The magnitude of the product 
of the $m$th row of $\MW_1$ and $\vx_j$
is proportional to the product of their Euclidean norms;
and the bias of the $m$th neuron -- the $m$th row of $\vb_1$ --
pushes its output (\ref{eq:neuronoutput}) to saturation.
A large positive bias pushes the output toward $1$
and a large negative bias pushes it to $0$.
Figure \ref{fig:DNNparameters}(a)
shows the Euclidean norms of all rows of $\MW_1$
for the classifier of the system of Fig. \ref{fig:DeSPerF_expt00},
sorted by descending norm.
Figure \ref{fig:DNNparameters}(b)
shows the bias of these neurons in the same order.
We immediately see from this that 
the inputs to almost half of the neurons in the first hidden layer
will have energies that are more than 20 dB below the neurons
receiving the most energy, and that they also display very small biases.
This suggests that about the half of the neurons in the first hidden layer
might be inconsequential to the system's behaviour.
In fact, when we neutralise the 250 neurons in the first hidden layer of DeSPerF-BALLROOM
having the smallest norm weights (by setting to zero the corresponding columns in $\MW_2$),
its FoM is identical to Fig. \ref{fig:DeSPerF_expt00}.
A possible explanation for this is that
the DNN has more parameters than are necessary
to map its input to its target. 

\subsection{Sensitivities and limitations}\label{sec:sensitivities}
\begin{figure}[t]
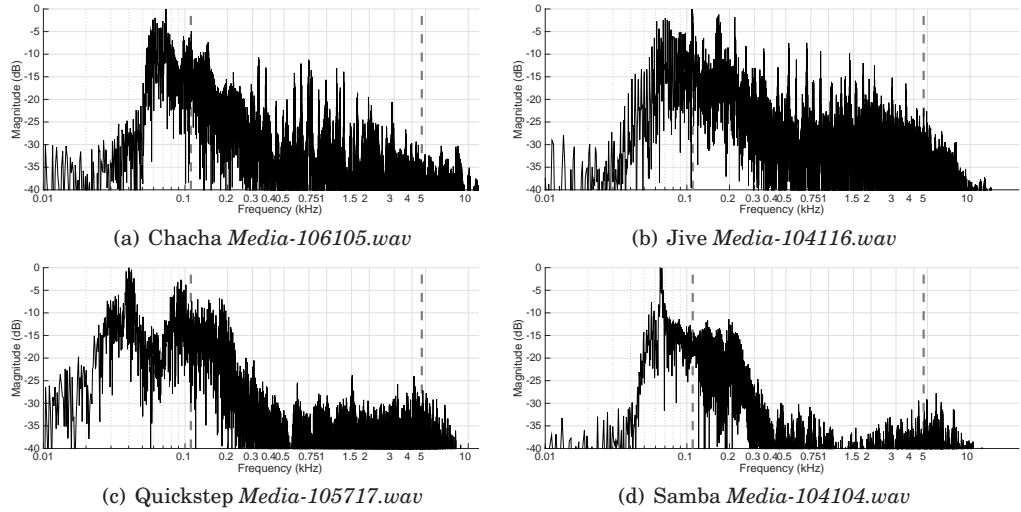

\centering
\subfigure[Chacha {\em Media-106105.wav}]{
\includegraphics[width=2.65in]{spectra_Chacha_Media-106105_ex.eps}}\hspace{-0.1in}
\subfigure[Jive {\em Media-104116.wav}]{
\includegraphics[width=2.65in]{spectra_Jive_Media-104116_ex.eps}}
\subfigure[Quickstep {\em Media-105717.wav}]{
\includegraphics[width=2.65in]{spectra_Quickstep_Media-105717_ex.eps}}\hspace{-0.1in}
\subfigure[Samba {\em Media-104104.wav}]{
\includegraphics[width=2.65in]{spectra_Samba_Media-104104_ex.eps}}
\caption{Magnitude spectra of two bars extracted from recordings in {\em BALLROOM}.
Dashed lines show bandwidth of filterbank in Fig. \ref{fig:SPerFMFCC}.}
\label{fig:samplespectra}
\end{figure}

From our analyses of the components of DeSPerF-based systems, 
we can infer the sensitivities and limitations 
of its feature extraction with respect to mapping $\UniverseRecording$
to $\UniverseSemanticFeature$, and its classification 
mapping $\UniverseSemanticFeature$ to $\UniverseSemantic$.
All of these limitations naturally restrict 
the $\UniverseMusic$, $\UniverseRecording$
and success criteria of a use case to which the system can be applied.
First, the MFCC filterbank (Fig. \ref{fig:SPerFMFCC}) means
this mapping is independent of any information outside of
the frequency band $[0.110, 4.43]$ kHz.
This could exclude most of the energy of bass drum kicks,
cymbal hits and crashes, shakers, and so on.
Figure \ref{fig:samplespectra} shows for segments of
four recordings from {\em BALLROOM} that a large amount of energy 
can exist outside this band.
If the information relevant for solving 
a problem of music content analysis is outside this band,
then a DeSPerF-based system may not be successful.

Second, since the segment modulation sonograms (\ref{eq:segmodsonogram}) 
consist of only the first 13 MFCCs,
their bandwidth is restricted to $[0, 0.093)$ cycles per semitone,
with a cepstral analysis resolution of not less than $10.7$ semitones.\footnote{
The period of the $k$th DCT function is $128/k$ semitones.}
Spectral structures smaller than about $11$ semitones
will thus not be present in a segment modulation sonogram.
If the information relevant for solving 
a problem of music content analysis is contained
only in spectral structures smaller than about 11 semitones (e.g., harmonic relationships of partials),
then a DeSPerF-based system may not be successful.

Third, the computation of the mean distance between 
lagged modulation sonograms (\ref{eq:meandistancelaggedmodsonograms})
destroys the quefrency information in the modulation sonograms.
In other words, there exist numerous modulation sonograms (\ref{eq:segmodsonogram}) 
that will produce the same mean distance sequence (\ref{eq:meandistancelaggedmodsonograms}).
This implies that SPerF (\ref{eq:SPerF}) are to a large extent 
invariant to timbre and pitch, and thus DeSPerF-based systems should 
not be sensitive to timbre and pitch, as long as the ``important information''
remains in the frequency band $[0.110, 4.43]$ kHz mentioned above.
This again restricts the kinds of problems of music content analysis
for which a DeSPerF-based system could be successful.

Fourth, since the mean distance between 
lagged modulation sonograms (\ref{eq:meandistancelaggedmodsonograms})
is uniformly sampled at a rate of $1/\durFrameSkip = 200$ Hz,
the frequency of repetition that can be represented in a SPerF (\ref{eq:SPerF})
is limited to the bandwidth $[0,100]$ Hz.
Furthermore, all repetitions at higher frequencies will be aliased to that band.
From our analysis of the front end of the DNN,
we see from Fig. \ref{fig:DNNweights}(b)
that DeSPerF-BALLROOM
is most sensitive to modulations in SPerF below 10 Hz.
In fact, the FoM of DeSPerF-BALLROOM in Fig. \ref{fig:DeSPerF_expt00} does not change
when we filter all input SPerF
with a zero-phase lowpass filter having a -3dB frequency of 10.3 Hz.
This implies that DeSPerF-BALLROOM
is not sensitive to SPerF modulations above 10 Hz,
which entails periods in SPerF of 100 ms or more.
Hence, DeSPerF-BALLROOM may have little sensitivity to periodicities in SPerF
that are shorter than 100 ms.

Finally, since each segment of a recording $r$ is of duration $\durSegment = 10$ s
for the system of Fig. \ref{fig:DeSPerF_expt00},
then a SPerF can only contain events repeating within that duration.
Since the largest lag considered is $\numframes\durFrameSkip = 4$ s,
this limits the duration of the periodic structures a SPerF can capture. 
For instance, if a periodic pattern of interest is of duration of one bar of music,
then a SPerF may only describe it if it repeats at least twice within 4 s.
For two consecutive repetitions, this implies that the
tempo must be greater than 120 BPM for a 4/4 time signature,
90 BPM for 3/4, and 180 BPM for 6/8.
If a repeated rhythm occurs over two bars,
then a SPerF may only contain it if at least four bars
occur within 4 s, or as long as the tempo is greater than 240 BPM for a 4/4 time signature,
180 BPM for 3/4, and 360 BPM for 6/8.

\subsection{Conclusion}
We have now dissected the system in Fig. \ref{fig:DeSPerF_expt00}.
We know that the DeSPerF-based systems are sensitive to temporal events
that repeat within a specific frequency band and particular time window.
This limits what DeSPerF-BALLROOM can be using to produce the 
FoM in Fig. \ref{fig:DeSPerF_expt00}.
For instance, because of its lack of spectral resolution,
it cannot be using melodies or harmonies 
to recognise elements of $\UniverseSemantic$.
Because it marginalises the quefrency information
it cannot be discriminating based on instrumentation.
It seems like the only knowledge a DeSPerF-based system
can be using must be temporal in nature within a 10-second window.
Before we can go further,
we must develop an understanding of how DeSPerF-BALLROOM
was trained and tested, and thus what Fig. \ref{fig:DeSPerF_expt00} might mean.
In the next section, we analyse the teaching and testing materials
used to produce DeSPerF-BALLROOM, 
and its FoM in Fig. \ref{fig:DeSPerF_expt00}.

\section{The Materials of Teaching and Testing}\label{sec:BALLROOM}
What is in the benchmark dataset {\em BALLROOM}?
What problem does it pose?
What is the task to reproduce its ground truth?
What is the goal or hope of training a music content analysis system with it? 
We now analyse the {\em BALLROOM} dataset used to train and test
DeSPerF-BALLROOM,
and how it has been used to teach and test 
other music content analysis systems.

\subsection{The contents and use of the {\em BALLROOM} dataset}
The dataset {\em BALLROOM}\footnote{Downloadable from \url{http://mtg.upf.edu/ismir2004/contest/tempoContest/node5.html}} 
consists of 698 half-minute music audio recordings
downloaded in Real Audio format around 2004 from an on-line 
resource about Standard and Latin ballroom dancing \cite{Dixon2004}.
Each excerpt comes from the ``beginning'' of a music track,
presumably ripped from a CD by an expert involved with the website.
\citeN{Dixon2004} call the labels of the dataset
both ``style'' and ``genre,'' and allude to each excerpt
being ``reliably labelled'' in one of eight ways.
Table \ref{tab:BALLROOMdataset} shows the distribution of the number of excerpts
over the labels of {\em BALLROOM} 
(we combine excerpts labeled ``Viennese Waltz'' and ``Waltz'' into one), 
as well as the 70/30 distribution of recordings 
we used for training and testing DeSPerF-BALLROOM in Fig. \ref{fig:DeSPerF_expt00}.

\begin{table}
\tbl{{\em BALLROOM} train/test partition used to train and test DeSPerF-BALLROOM
in Fig. \ref{fig:DeSPerF_expt00}}{
\begin{tabular}{|r|c|c|c|}
Label & Train & Test & Totals\\ 
& No. (\%) & No. (\%) & No. (\%) \\\hline
ChaCha & 78 (15.85) & 33 (16.02)  & 111 (15.90) \\ 
Jive & 42 (8.54) & 18 (8.74)  & 60 (8.60) \\ 
Quickstep & 58 (11.79) & 24 (11.65)  & 82 (11.75) \\ 
Rumba & 69 (14.02) & 29 (14.08)  & 98 (14.04) \\ 
Samba & 61 (12.40) & 25 (12.14)  & 86 (12.32) \\ 
Tango & 61 (12.40) & 25 (12.14)  & 86 (12.32) \\ 
Waltz & 123 (25.00) & 52 (25.24)  & 175 (25.07) \\ 
\hline Total & 492 (70.49) & 206 (29.51) & 698 (100)\\ 
\hline\end{tabular}}
\label{tab:BALLROOMdataset}
\begin{tabnote}\end{tabnote}\end{table}

\begin{table}%
\tbl{All appearances of {\em BALLROOM} \cite{Dixon2004} 
in published classification experiments,
along with the highest reported accuracy (normalised or not).}{%
\begin{tabular}{|r|l|}
\hline
 & Highest Acc. \\
Reference &  Reported (\%) \\
\hline
\citeN{Dixon2004} & 96  \\
\citeN{Gouyon2004} & 90.1  \\
ISMIR2004 & 82 \\
\citeN{Gouyon2004b},\citeN{Gouyon2005} & 82.1  \\
\citeN{Lidy2005} & 84.24  \\
\citeN{Peeters2005} & 90.4  \\
\citeN{Flexer2006} & 66.9  \\
\citeN{Lidy2006} & 82  \\
\citeN{Lidy2007} & 90.4 \\
\citeN{Lidy2008} & 90.0  \\
\citeN{Holzapfel2008b} & 85.5  \\
\citeN{Holzapfel2009} & 86.9 \\
\citeN{Pohle2009} & 89.2 \\
\citeN{Lidy2010b} & 87.97  \\
\citeN{Mayer2010c} & 88  \\
\citeN{Seyerlehner2010},\citeN{Seyerlehner2010b} & $\sim$ 90  \\
\citeN{Peeters2011} & 96.1 \\
\citeN{Tsunoo2011} & 77.2 \\ 
\citeN{Schindler2012b} & 67.3 \\
\citeN{Pikrakis2013} & $\sim$ 85  \\
\citeN{Sturm2014} & 88.8  \\
\hline
\end{tabular}}
\label{tab:BALLROOMresults}
\begin{tabnote}%
\end{tabnote}%
\end{table}%

Thus far, {\em BALLROOM} has appeared in the evaluations of
at least 24 conference papers, journal articles, and PhD dissertations
\cite{Dixon2004,Flexer2006,Gouyon2004,Gouyon2004b,Gouyon2005,Holzapfel2008b,Holzapfel2009,Lidy2005,Lidy2006,Lidy2007,Lidy2008,Lidy2010b,Mayer2010c,Peeters2005,Peeters2011,Pikrakis2013,Pohle2009,Schindler2012b,Schluter2011,Schnitzer2011,Schnitzer2012,Seyerlehner2010,Seyerlehner2010b,Seyerlehner2012,Tsunoo2011}.
Twenty of these works use it in the experimental design {\em Classify} \cite{Sturm2014d},
which is the comparison of ground truth to the output of a music content analysis system.
Table \ref{tab:BALLROOMresults} shows the highest accuracies
reported in the publications using {\em BALLROOM} this way.
Four others \cite{Schluter2011,Schnitzer2011,Schnitzer2012,Seyerlehner2012}
use {\em BALLROOM} in the experimental design {\em Retrieve} \cite{Sturm2014d},
which is the task of retrieving music signals from the training set given a query.
The dataset was also used for the Rhythm Classification Train-test Task
of ISMIR2004,\footnote{\url{http://mtg.upf.edu/ismir2004/contest/rhythmContest/}}
and so sometimes appears as {\em ISMIRrhythm}.

\subsection{Some tasks posed by the {\em BALLROOM} dataset}\label{sec:BALLROOMproblems}
\begin{figure}[t]
\centering
\subfigure[Tango excerpt {\em Albums-Ballroom Classics4-07} and {\em Media-105705}]{\includegraphics[height=0.4in]{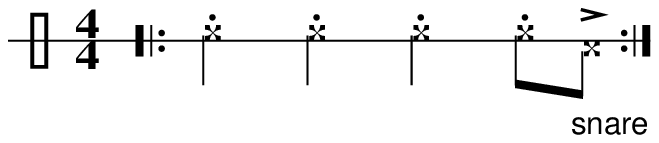}\label{fig:patterns_BALLROOM_tango}}
\subfigure[Samba excerpt {\em Albums-Latin Jam-06} and {\em Media-103901}]{\includegraphics[height=0.6in]{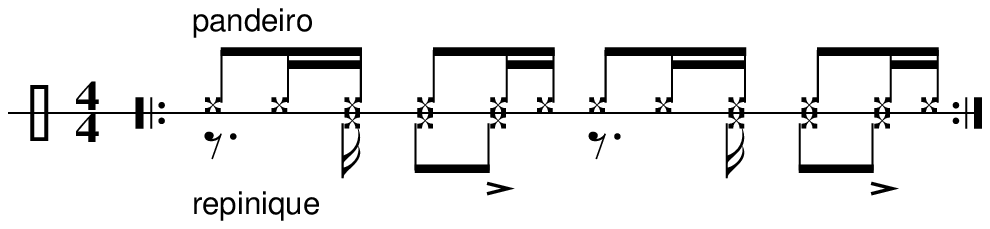}\label{fig:patterns_BALLROOM_samba}}
\subfigure[Cha cha excerpts {\em Albums-Latin Jam3-02} and
{\em Albums-Latin Jam4-06}]{\includegraphics[height=0.8in]{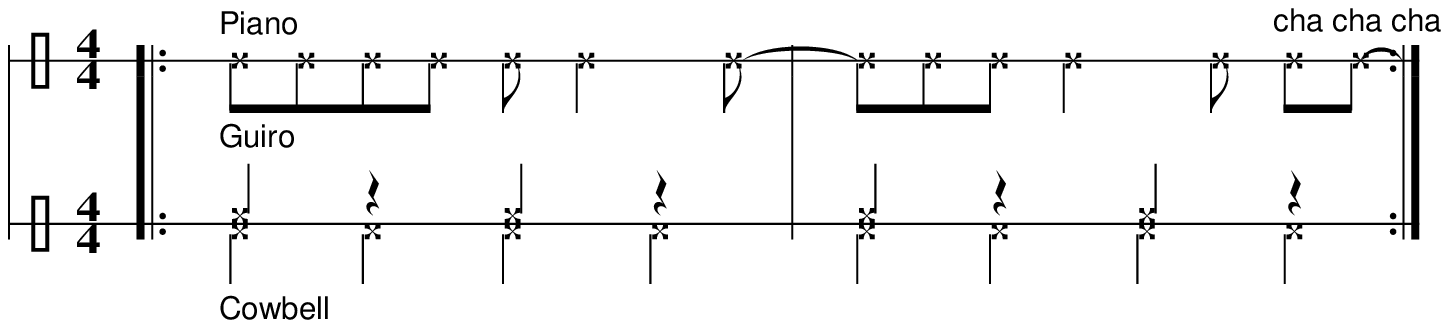}\label{fig:patterns_BALLROOM_chacha}}
\subfigure[Rumba excerpt {\em Media-105614}]{\includegraphics[height=0.6in]{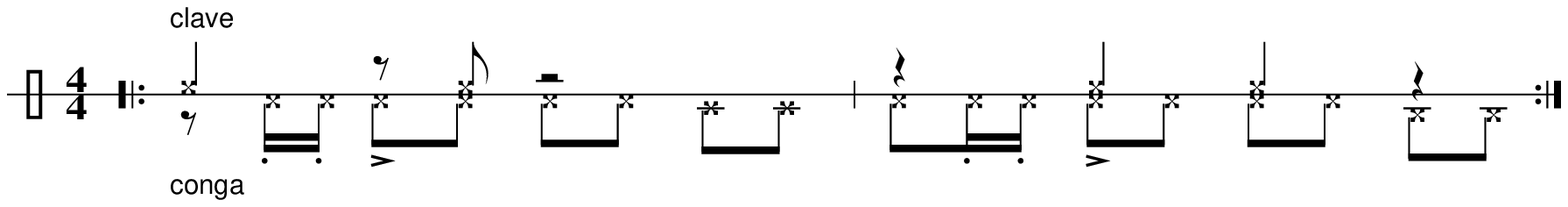}\label{fig:patterns_BALLROOM_rumba}}
\subfigure[Jive excerpt {\em Albums-Fire-12}]{\includegraphics[height=1in]{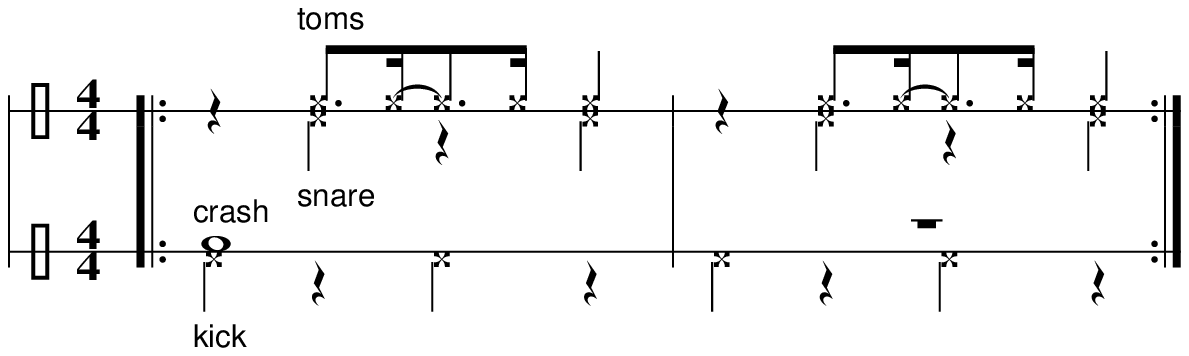}\label{fig:patterns_BALLROOM_jive}}
\subfigure[Quickstep excerpt {\em Albums-AnaBelen Veneo-11}]{\includegraphics[height=1in]{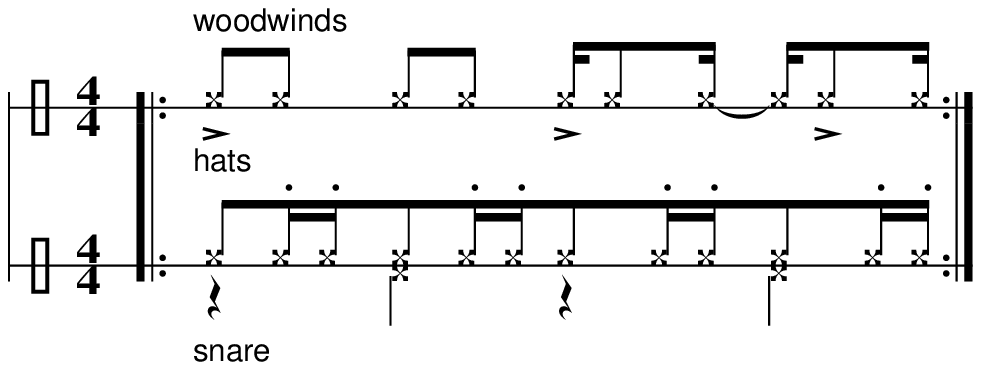}\label{fig:patterns_BALLROOM_quickstep}}
\caption{Some of the characteristic patterns found in excerpts of {\em BALLROOM}.}
\label{fig:patterns_BALLROOM}
\end{figure}

\citeN{Dixon2004} and \citeN{Gouyon2004} pose one task of {\em BALLROOM} as to extract
and learn ``repetitive rhythmic patterns'' from recorded music audio indicating the correct label.
Motivating their work and the creation of the dataset,
\citeN{Dixon2004} propose the hypothesis:
``rhythmic patterns are not randomly distributed amongst musical genres, 
but rather they are indicative of a genre.''
While ``rhythm'' is an extraordinarily difficult thing to define \cite{Gouyon2005},
examples illuminate what \citeN{Dixon2004} and \citeN{Gouyon2004} intend.
For instance, they give one ``rhythmic pattern'' typical of Cha cha and Rumba 
as one bar of three crochets followed by two quavers.
Auditioning the Cha cha recordings
reveals that this pattern does appear but 
that it can be quite difficult to hear through the instrumentation.
In fact, this pattern is also apparent in many of 
the Tango recordings (notated in Fig. \ref{fig:patterns_BALLROOM_tango}).
We find that major differences between recordings 
of the two labels are instrumentation, 
the use of accents, and syncopated accompaniment.
It should be noted that much of the ``rhythmic information'' 
in excerpts of several labels of {\em BALLROOM}
is contributed by instruments other than percussion, 
such as the piano and guitar in Cha cha, Rumba, Jive, Quickstep, and Tango;
brass sections, woodwinds and electric guitar in Jive and Quickstep;
and vocals and orchestra in Waltz.

Figure \ref{fig:patterns_BALLROOM} shows 
examples of the rhythmic patterns appearing in {\em BALLROOM}.
By ``rhythmic pattern'' we mean a combination of 
metrical structure, and relative timing and accents
in a combination of voices.
Many Cha cha recordings feature a two bar pattern with a 
strong cowbell on every beat,
a guiro on one and three, 
and syncopated piano and/or brass
with notes held over the bars (notated in Fig. \ref{fig:patterns_BALLROOM_chacha}).
On the other hand, Rumba recordings 
sound much slower and sparser than those of Cha cha,
often featuring only guitar, clave, conga, shakers,
and the occasional chime glissando (Fig. \ref{fig:patterns_BALLROOM_rumba}).
Rhythmic patterns heard in Jive and Quickstep recordings
involve swung notes,
notated squarely in Fig. \ref{fig:patterns_BALLROOM_jive} and
Fig. \ref{fig:patterns_BALLROOM_quickstep}. 
We find no Waltz recordings to have duple or quadruple meter.

Even though this dataset was explicitly created
for the task of learning ``repetitive rhythmic patterns,''
it actually poses other tasks.
In fact, a music content analysis system
need not know one thing about rhythm to reproduce
the ground truth in {\em BALLROOM}.
One such task 
is the identification of instruments.
For instance, bandoneon only appears in Tango recordings. 
Jive and Quickstep recordings
often feature toms and brass,
but the latter also has woodwinds.
Rumba and Waltz recordings
feature string orchestra,
but the former also has chimes and conga.
Cha cha recordings often have 
piano, along with guiro and cowbell.
Finally, Samba recordings feature
instruments that do not occur in any other recordings,
such as pandeiro, repinique, whistles, and cuica.
Hence, a system completely naive to rhythm
could reproduce the ground truth of {\em BALLROOM}
just by recognising instruments.
This clearly solves a completely different problem from that posed by 
\citeN{Dixon2004} and \citeN{Gouyon2004}.
It is aligned more with the task posed by \citeN{Lidy2008}:
``to extract suitable features from a benchmark music collection 
and to classify the pieces of music into a given list of genres.''

%
%
%
%
%
%
%

\begin{figure}[t]
\centering
\includegraphics[width=4in]{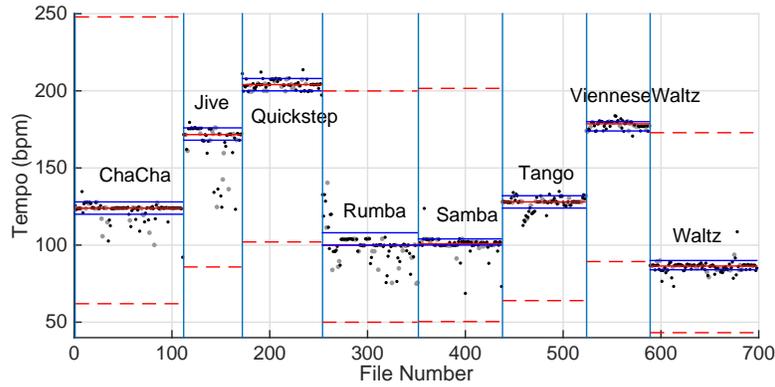}
\caption{Distribution of the tempi of recordings (dots) in BALLROOM, 
assembled from onset data of \citeN{Krebs2013}.
For each label: red solid line is median tempo;
red dotted lines are half and double media tempo;
upper and lower blue lines are official tempos
for acceptable dance competition music by the \citeN{WSDF2014}
(see Table \ref{tab:BALLROOMregulations});
black dots are recordings in training dataset
used to build DeSPerF-BALLROOM,
and grey dots are recordings in the test dataset
to compute its FoM in Fig. \ref{fig:DeSPerF_expt00}.
}
\label{fig:BALLROOM_tempo}
\end{figure}

\begin{figure}[t]
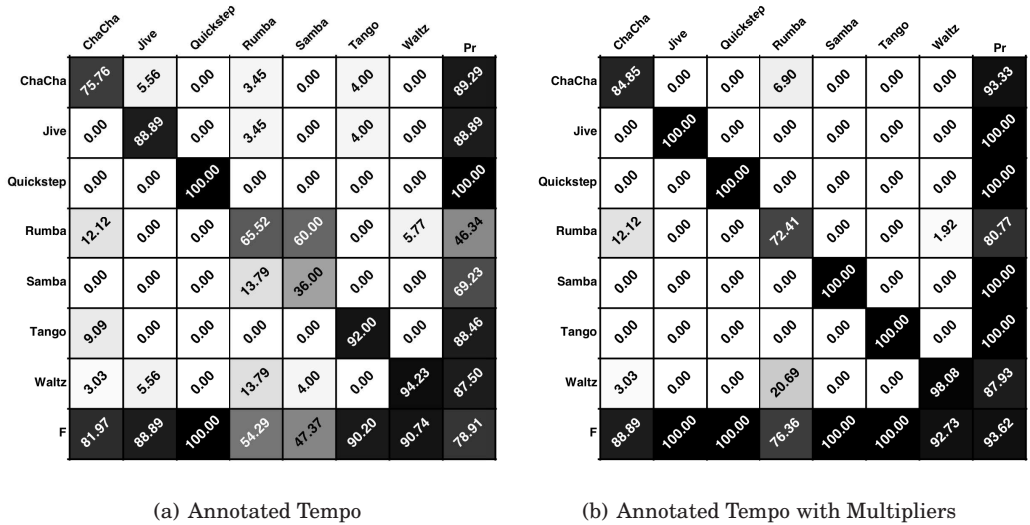

\centering
\subfigure[Annotated Tempo]{\hspace{-0.2in}
\includegraphics[width=2.8in]{BALLROOMtempo_1nn.eps}}\hspace{-0.1in}
\subfigure[Annotated Tempo with Multipliers]{
\includegraphics[width=2.8in]{BALLROOMtempo_1nn_wdoublings.eps}}
\caption{FoM of single nearest neighbour classifiers using just tempo
for classification of excerpts in {\em BALLROOM}.
Interpretation as in Fig. \ref{fig:DeSPerF_expt00}.}
\label{fig:BALLROOM_tempo_NN}
\end{figure}

There exists yet another way to reproduce the ground truth of {\em BALLROOM}.
Figure \ref{fig:BALLROOM_tempo} shows the distribution of tempi.
We immediately see a strong correlation between tempo and label.
This was also noted by \citeN{Gouyon2004}.
To illustrate the strength of this relationship,
we construct a music content analysis system using
simple nearest neighbour classification \cite{Hastie2009} with tempo alone.
Figure \ref{fig:BALLROOM_tempo_NN}(a) shows the FoM
of this system using the same training and testing partition of {\em BALLROOM}
as in Fig. \ref{fig:DeSPerF_expt00}.
Clearly, this system produces a significant amount of ground truth,
but suffers from a confusion predictable from Fig. \ref{fig:BALLROOM_tempo} --
which curiously does not appear in Fig. \ref{fig:DeSPerF_expt00}.
If we modify annotated tempi 
by the following factors:
Cha cha $\times 2$;
Jive $\times 0.5$; 
Quickstep $\times 0.5$; 
Rumba $\times 2$; 
Samba $\times 0.5$;
Tango $\times 1$;
and Waltz $\times 2$ (keeping Viennese Waltz the same),
then the new system
produces the FoM in Figure \ref{fig:BALLROOM_tempo_NN}(b).
Hence, ``teaching'' the system to ``tap its foot'' 
half as fast for some labels,
and twice as fast for others,
ends up reproducing a similar amount of ground truth
to DeSPerF-BALLROOM in Fig. \ref{fig:DeSPerF_expt00}.

While such a foot-tapping system can reproduce the labels of {\em BALLROOM},
the particular problem it is actually solving is not aligned with
that of detecting ``repetitive rhythmic patterns'' \cite{Dixon2004,Gouyon2004}.
The system of Fig. \ref{fig:BALLROOM_tempo_NN} 
is also not solving the problem posed by \citeN{Lidy2008} 
as long as ``genre'' is not so strongly characterised by tempo.
Of course, there are official tempos set by the \citeN{WSDF2014}
for music to be acceptable for dance competitions 
(see Fig. \ref{fig:BALLROOM_tempo} and Table \ref{tab:BALLROOMregulations}),
but arguably these rules are created to balance
skill and competition difficulty,
and are not derived from surveys of musical practice,
and certainly are not proscriptions 
for the composition and performance of music in these styles.
In fact, Fig. \ref{fig:BALLROOM_tempo} shows several 
{\em BALLROOM} recordings do not satisfy these criteria.

\begin{table}%
\tbl{Ballroom dance music tempo regulations 
of the \citeN{WSDF2014}.}{%
\begin{tabular}{|c|c|c|}
Dance Style & Tempo regulation & Scale factor \\ 
& bars/min (beats/min) & from mean tempo \\
\hline
Cha-Cha-Cha & 30 - 32 (120 - 128) & 0.969 - 1.033 \\
Jive & 42 - 44 (168 - 176) & 0.977 - 1.024 \\
Quickstep & 50 - 52 (200 - 208) & 0.981 - 1.020\\
Rumba & 25 - 27 (100 - 108) & 0.963 - 1.040 \\
Samba & 50 - 52 (100 - 104) & 0.981 - 1.020 \\
Tango & 31 - 33 (124 - 132) & 0.970 - 1.032 \\
Viennese Waltz & 58 - 60 (174 - 180) & 0.983 - 1.017\\
Waltz & 28 - 30 (84 - 90) & 0.967 - 1.036 \\
\hline
\end{tabular}}
\label{tab:BALLROOMregulations}
\begin{tabnote}%
\end{tabnote}%
\end{table}%

Reproducing the ground truth of {\em BALLROOM}
by performing any of the tasks above -- 
discrimination by ``rhythmic patterns,'' instrumentation, and/or tempo --
clearly involves using high level acoustic and musical characteristics.
However, there are yet other tasks that
a system might be performing 
to reproduce the ground truth of {\em BALLROOM},
and ones with no clear relationship to music listening.
For instance, if we use single nearest neighbour classification
with features composed of only the variance and mean of a SPerF, 
and the number of times it passes through $0.5$, 
then with majority voting this system obtains 
a classification accuracy of over $0.70$ --
far above that expected by random classification.
It is not clear what task this system is performing,
and how it relates to 
high-level acoustic and musical characteristics.
Hence, this fourth approach to 
reproducing the ground truth of {\em BALLROOM}
solves an entirely different problem
from the previous three: 
``to classify the music documents 
into a predetermined list of classes'' \cite{Lidy2005},
i.e., {\em by any means possible.}

\subsection{Conclusion}
Though the explicit and intended task of {\em BALLROOM}
is to recognise and discriminate between rhythmic patterns,
we see that there actually exists many other tasks a system
could be performing in reproducing the ground truth.
The common experimental approach in music content analysis 
research, i.e., that used to produce the FoM in Fig. \ref{fig:DeSPerF_expt00},
has no capacity to distinguish between any of them. 
Just as in the case
for the demonstrations of Clever Hans,
were a music content analysis system
actually recognising characteristic rhythms of
some of the labels of {\em BALLROOM},
its FoM might pale in comparison to that of
a system with no idea at all about rhythm (Fig. \ref{fig:BALLROOM_tempo_NN}).
Figure \ref{fig:DeSPerF_expt00} gives no evidence at all 
for claims that DeSPerF-BALLROOM
is identifying waltz by recognising its characteristic rhythmic patterns,
tempo, instrumentation, and/or any other factor.
From our analysis of DeSPerF-based systems,
however, we can rule out instrument recognition
since such knowledge is outside its purview.
Nonetheless, what exact ask DeSPerF-BALLROOM
is performing, the {\em cause} of Fig. \ref{fig:DeSPerF_expt00}, remains to be seen.
The experiments in the next section shed light on this.


%
\section{Seeking the ``Horse'' Inside the Music Content Analysis System}\label{sec:experiments}
It is obvious 
that DeSPerF-BALLROOM knows {\em something} about the 
recordings in {\em BALLROOM};
otherwise its FoM in Fig. \ref{fig:DeSPerF_expt00} would not be so significantly
different from chance.
As discussed in the previous section,
this might be due to the system 
performing any of a number of tasks,
whether by identifying rhythms, detecting tempo, 
or using the distributions of statistics with completely obscured relationships
to music content.
In this section, we describe several experiments
designed to explain Fig. \ref{fig:DeSPerF_expt00}.


\subsection{Experiment 1: The nature of the cues}
We first seek the nature of the cues
used by DeSPerF-BALLROOM to reproduce the ground truth.
We watch how its behaviour changes when we 
modify the input along two orthogonal dimensions:
frequency and time.
We transform recordings of the test dataset by 
pitch-preserving time stretching,
and time-preserving pitch shifting.\footnote{We use the rubberband library
to achieve these transformations with minimal change in recording quality.
We have auditioned several of the transformations to confirm.}
We seek the minimum scalings to make the system obtain a perfect classification accuracy,
or one consistent with random classification (14.3\%).
To ``inflate'' the FoM,
we take each test recording for which DeSPerF-BALLROOM is incorrect
and transform it using a scale
that increments by $0.01$ until the system is no longer incorrect.
To ``deflate'' the FoM,
we take each test recording for which DeSPerF-BALLROOM is correct
and transform it using a scale
that increments by $0.01$ until it is no longer correct.
A pitch-preserving time stretching of scale $1.05$
increases the recording duration by $5\%$,
or decreases the tempo of the music in the recording 
(if it has a tempo) by $5\%$.
A time-preserving pitch shifting of scale $1.05$
increases all pitches in a recording by $5\%$.

\begin{figure}[t]
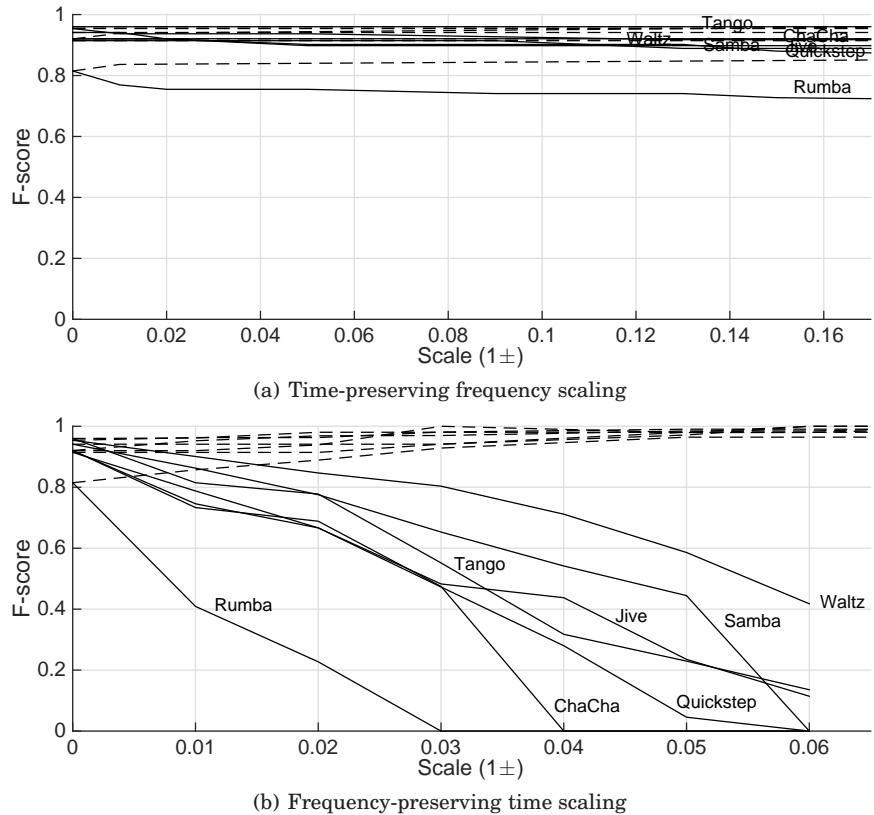

\centering
\subfigure[Time-preserving frequency scaling]{
\includegraphics[width=4.5in]{FoM_DeSPerF_BALLROOM_freq_F-score.eps}}
\subfigure[Frequency-preserving time scaling]{
\includegraphics[width=4.5in]{FoM_DeSPerF_BALLROOM_time_F-score.eps}}
\caption{Changes to F-score of each label of {\em BALLROOM}
as a function of the scaling of a transformation.
Solid lines: deflation procedure.
Dashed lines: inflation procedure. 
Note the difference in scales on the x-axis.}
\label{fig:BALLROOM_expt01}
\end{figure}

\begin{figure}[t]
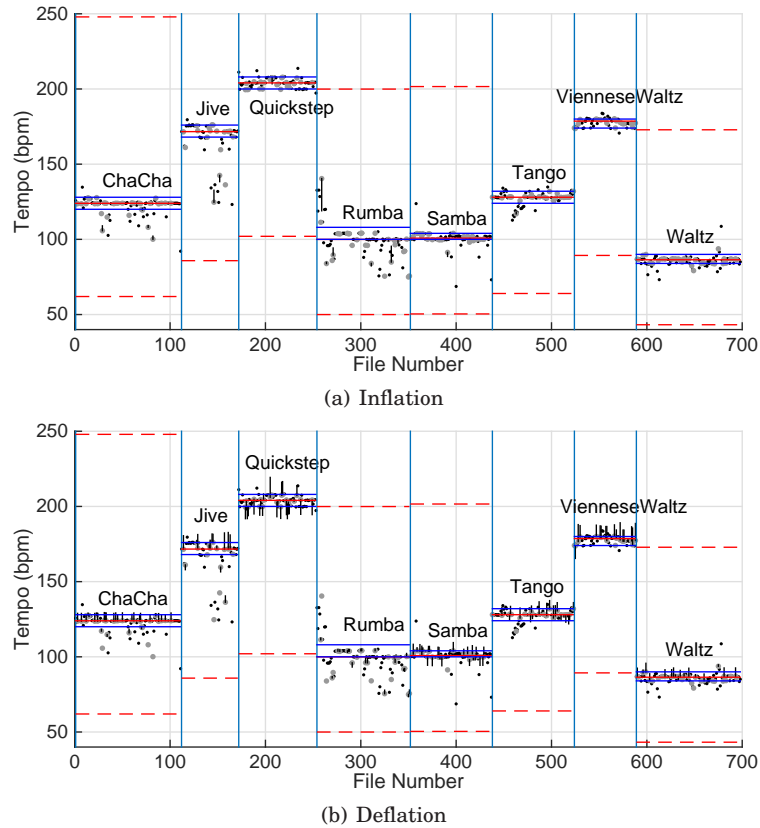

\centering
\subfigure[Inflation]{
\includegraphics[width=3.9in]{tempo_BALLROOM_INFLATE.eps}}
\subfigure[Deflation]{
\includegraphics[width=3.9in]{tempo_BALLROOM_DEFLATE.eps}}
\caption{Vertical lines point from original tempo of a 
{\em BALLROOM} test recording (grey dot)
to its tempo after transformation 
by pitch-preserving time stretching of at most $6$\%.
Interpretation as in Fig. \ref{fig:BALLROOM_expt01}.}
\label{fig:BALLROOM_expt01_tempo}
\end{figure}

Figure \ref{fig:BALLROOM_expt01} shows the results.
As expected from our analysis in Section \ref{sec:sensitivities}, 
time-preserving pitch shifting of the test recordings
has little effect on the FoM,
even up to changes of $\pm 16\%$.
In stark contrast is the effect of pitch-preserving time stretching,
where the F-score of DeSPerF-BALLROOM in each label
quickly decays for scales of at most $\pm 5\%$.
That scale is equivalent to lengthening or shortening
a 30 s recording by only 1.5 s.
Figure \ref{fig:BALLROOM_expt01_tempo} shows the 
new tempi of the test recordings after these procedures, 
i.e., when the normalised classification accuracy
is either perfect or no better than random.
We see in most cases that the tempo changes are very small.
The tempi of the 16 test recordings initially classified wrong
move toward the median tempo of each class.
Figure \ref{fig:BALLROOM_expt01_tempo}(b)
shows that the opposite occurs in deflation 
for the 190 test recordings initially classified correctly.

The effects of these transformations  
clearly show that the nature of the cues DeSPerF-BALLROOM
uses to reproduce ground truth is temporal,
and that its performance is completely disrupted by
minor changes in music tempo.
The mean tempo change of the 
12 {\em BALLROOM} Cha cha excerpts
in Fig. \ref{fig:BALLROOM_expt01_tempo}(b)
is an increase of $3.7$ BPM, 
which situate all of them on the cusp of
the Cha cha cha competition dance tempo regulation
(Table \ref{tab:BALLROOMregulations}).
Most of these transformed recordings
are then classified by the system as Tango.
In light of this, it is problematic to claim, e.g.,
DeSPerF-BALLROOM has such a high precision in 
identifying Cha cha (Fig. \ref{fig:DeSPerF_expt00})
because its internal model of Cha cha embodies ``typical rhythmic patterns''
of cha cha.
Something else is at play.

\subsection{Experiment 2: System dependence on the rate of onsets}
The results of the previous experiment 
suggest that if the internal models of DeSPerF-BALLROOM
have anything to do with rhythmic patterns,
they are such that minor changes to tempo
produce major confusion.
We cannot say that the specific temporal cue used by DeSPerF-BALLROOM is tempo --
however that is defined -- alone or in combination with other characteristics,
such as accent and meter.
Indeed, comparing Fig. \ref{fig:DeSPerF_expt00} with Fig. \ref{fig:BALLROOM_tempo_NN}
motivates the hypothesis that DeSPerF-BALLROOM is using tempo,
but reduces confusions by halving or doubling tempo based on something else.
In this experiment, we investigate the inclinations of
DeSPerF-BALLROOM to classify synthetic recordings exhibiting 
unambiguous onset rates.
We synthesise each recording in the following manner.
We generate one realisation of a white noise burst with duration 68 ms,
windowed by half of a Hann window (attack and smooth decay).
The burst has a bandwidth covering
the bandwidth of the filterbank 
in DeSPerF-BALLROOM (Section \ref{sec:features}).
We synthesise a recording by
repeating the same burst (no change in its amplitude) at a regular periodic interval
(reciprocal of onset rate),
and finally add white Gaussian noise with a power of 60 dB SNR
(to avoid producing features that are not numbers).
We create 200 recordings in total, 
with onset rates logarithmically spaced
from 50 to 260 onsets per minute.
Finally, we record the output of the system for each recording,
as well as the mean DNN output posterior (\ref{eq:DNNoutput}) over all segments.




\begin{figure}[t]
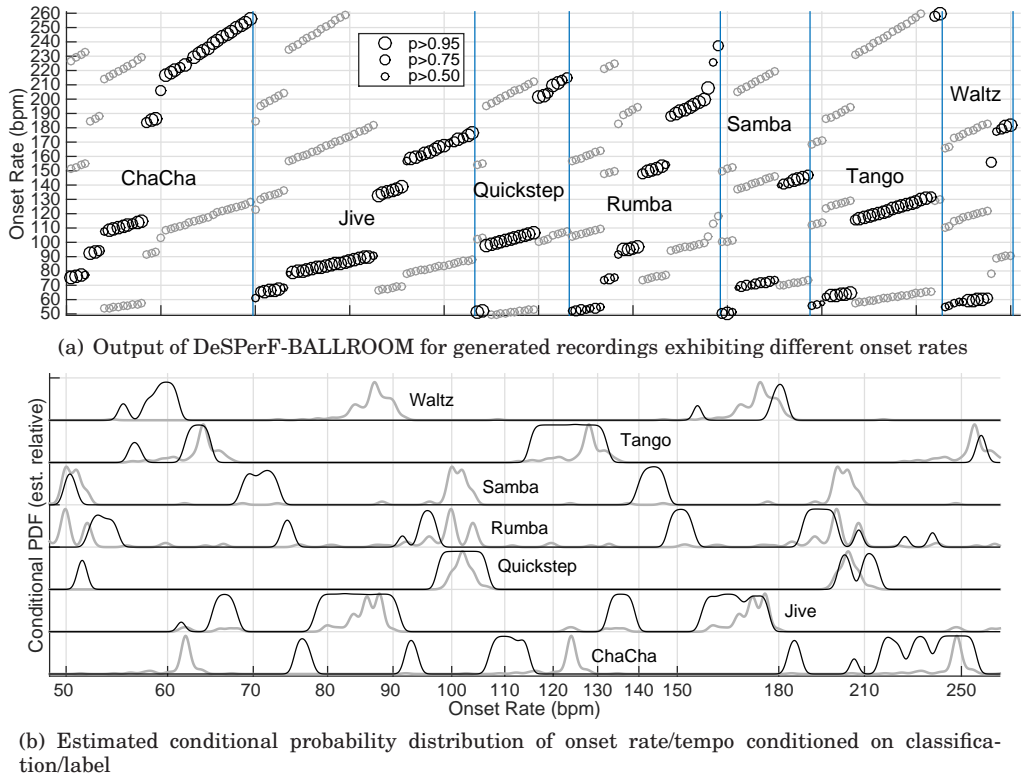

\centering
\subfigure[Output of DeSPerF-BALLROOM for generated recordings exhibiting different onset rates]{
\includegraphics[width=5.3in]{DeSPerF_BALLROOM_random_tempoclassifications.eps}}
\subfigure[Estimated conditional probability distribution
of onset rate/tempo conditioned on classification/label]{
\includegraphics[width=5.1in]{DeSPerF_BALLROOM_tempoclassifications_pdf.eps}}
\caption{Results from testing DeSPerF-BALLROOM
using synthetic recordings having different onset rates.
(a) A black circle is a recording with an onset rate (y-axis),
classified by DeSPerF-BALLROOM with mean posterior $p$ (legend).
We plot the halves and doubles of the onsets as well
as grey circles of the same size.
(b) Parzen window estimate of probability distributions of onset rate
conditioned on system output (black),
and tempo of training excerpts conditioned on label (grey)
with halving and doubling.}
\label{fig:DeSPerF_BALLROOM_expt02}
\end{figure}

Figure \ref{fig:DeSPerF_BALLROOM_expt02} shows the results of this experiment.
Each black circle in Fig. \ref{fig:DeSPerF_BALLROOM_expt02}(a)
represents a recording with some onset rate (y-axis),
classified by the system in some way 
(grouped in classes and ordered by increasing onset rate)
with a mean posterior $p$ (size of circle).
Figure \ref{fig:DeSPerF_BALLROOM_expt02}(b) shows
an estimates of the conditional distributions of onset rate
given the classification by using Parzen windowing with the posteriors as weights.
We also show the estimate of the conditional distribution of tempo 
given the {\em BALLROOM} label from the training data,
and include a halving and doubling of tempo (gray). 
We can clearly see ranges of onset rates
to which the system responds confidently
in its mapping.
Comparing the two conditional distributions,
we see some that align very well.
All octaves of the tempo of Jive, Quickstep and Tango
overlap the ranges of onsets that are 
confidently so classified by DeSPerF-BALLROOM.
For Samba, however, only the distribution of half the tempo
overlaps the Samba-classified synthetic recordings at low onset rates;
for Cha cha and Rumba, it is the distributions of double the tempo
that overlap the Cha cha- or Rumba-classified synthetic recordings at high onset rates.
These are some of the tempo multiples used to produce
the FoM in Fig. \ref{fig:BALLROOM_tempo_NN}(b)
by single nearest neighbour classification.
These results point to the hypothesis that
DeSPerF-BALLROOM
is using a cue to ``hear'' an input recording
at a ``tempo'' that best separates it from the other labels.
Of interest is whether that cue
has to do with meter and/or rhythm,
and how the system's internal models 
reflect high level attributes of
the styles in {\em BALLROOM}.
We explore these in the next three experiments.



\subsection{Experiment 3: System output dependence on the rate of onsets and periodic stresses}
In this experiment, we watch how the system's behaviour changes
when the input exhibits repeating structures that have a period 
encompassing several onsets.
We perform this experiment in the same manner as the previous one.
We synthesise each recording in the same way,
but stress every second, third or fourth repetition of the white noise burst.
We create a stress in two different ways.
In the first, 
each stressed onset has an amplitude four times that of an unstressed onset.
In the second, all unstressed onsets are produced by 
a highpass filtering of the white noise burst (passband frequency 1 kHz).
We create 200 recordings in total for each of the stress periods,
and each kind of stress, with onset rates logarithmically spaced
from 50 to 260 onsets per minute.
Finally, we record the output of the system for each recording,
as well as the mean DNN output posterior (\ref{eq:DNNoutput}) for all segments.

\begin{figure}[t]
\centering
\subfigure[System output for generated recordings exhibiting different onset rates and stress periods]{
\includegraphics[width=5.2in]{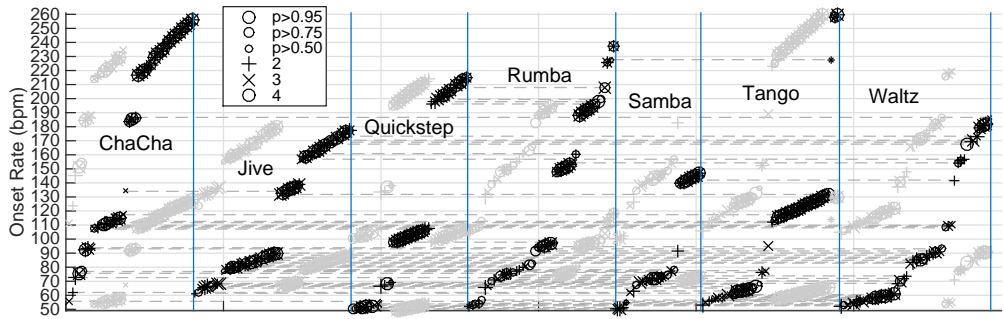}}
\subfigure[Estimated conditional probability distribution
of onset rate/tempo conditioned on classification/label and stress period]{
\includegraphics[width=5.1in]{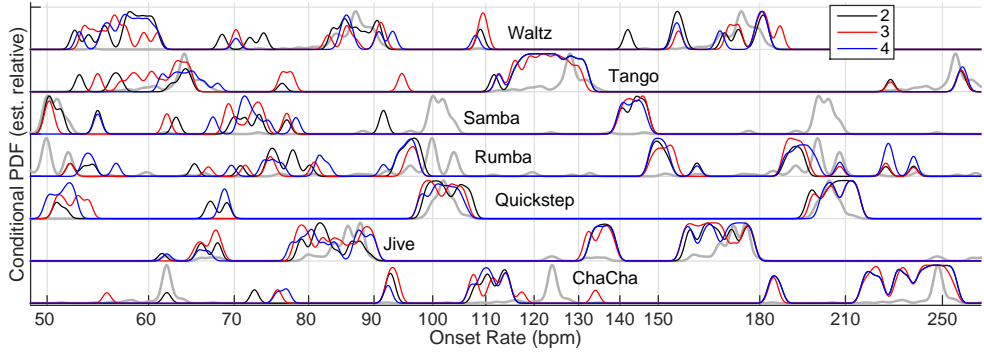}}
\caption{Results from testing DeSPerF-BALLROOM
using recordings generated with different onset rates and stress periods (legend).
Compare with Fig. \ref{fig:DeSPerF_BALLROOM_expt02}.
Horizontal dashed lines signify changes in class across stress period.
As in Fig. \ref{fig:DeSPerF_BALLROOM_expt02}(a),
we show halving and doubling of onset rates.}
\label{fig:DeSPerF_BALLROOM_expt03}
\end{figure}



\begin{figure}[t]
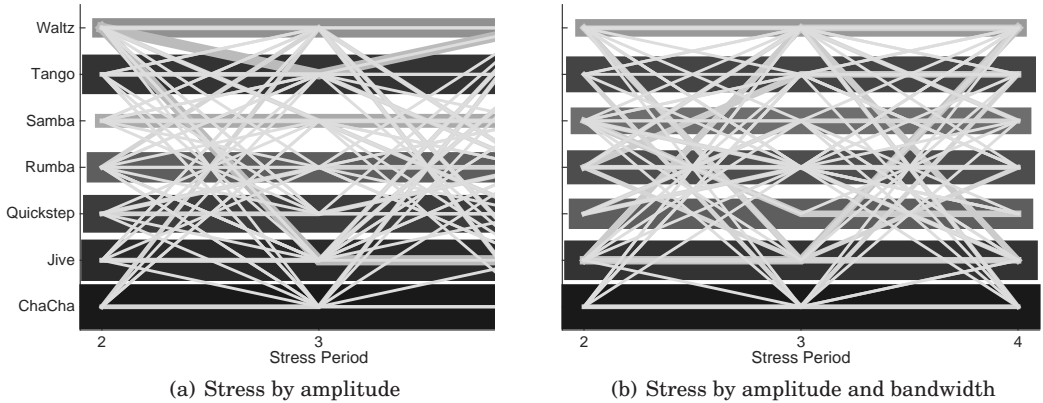

\centering
\subfigure[Stress by amplitude]{\includegraphics[height=1.9in]{DeSPerF_BALLROOM_meter_classifications.eps}}\hspace{-0.02in}
\subfigure[Stress by amplitude and bandwidth]{\includegraphics[height=1.9in]{DeSPerF_BALLROOM_metertimbre_classifications.eps}}
\caption{Dependency of system output (y-axis) on stress period
for two different kinds of stresses across all onset rates tested.
The weight of a line shows the proportion observed 
of a specific transition in classification for recordings generated with the same onset rate.
The transition pattern observed most in both cases (24 times)
is Cha cha (stress period 2), Cha cha (3), Cha cha (4).
}
\label{fig:DeSPerF_BALLROOM_expt03_dep}
\end{figure}

Figure \ref{fig:DeSPerF_BALLROOM_expt03} 
shows results quite similar to the previous experiment.
The results of both stress kinds are nearly the same,
so we only not show one of them.
The dashed horizontal lines in Fig. \ref{fig:DeSPerF_BALLROOM_expt03}(a)
show some classifications of recordings with the same onset rate 
are different across the stress periods we test.
Figure \ref{fig:DeSPerF_BALLROOM_expt03}(b)
shows the appearance of density in the conditional probability distribution of 
the onset rate in Waltz around the tempo distribution
observed in the training dataset of label Waltz (80-90 BPM),
which is not apparent in Fig. \ref{fig:DeSPerF_BALLROOM_expt02}(b).
Could these changes be due to the system preferring Waltz
for a recordings exhibiting a stress period of 3?
Figure \ref{fig:DeSPerF_BALLROOM_expt03_dep} shows
this to not be the case.
We see no clear indication that DeSPerF-BALLROOM
favours particular classes for each stress period independent of 
the onset rate for the different kinds of stresses.
For instance, we see no strong inclination of DeSPerF-BALLROOM 
to classify recordings with a stress period of 3 as Waltz.
Most classifications are the same across the stress periods.

\subsection{Experiment 4: Manipulation of the tempo}
The previous experiments clearly show the inclination of
DeSPerF-BALLROOM to classify in confident ways
recordings exhibiting specific onset rates
independent of repeated structures of longer periods.
This leads to the prediction that any input recording can be 
time-stretched to elicit any desired response from the system,
e.g., we can make the system choose ``Tango'' by
time stretching any input recording
to have a tempo of 130 BPM.
To test this prediction, we first observe how the system output changes when we
apply frequency-preserving time stretching to the entire {\em BALLROOM} test dataset
with scales from $0.5$ to $1.5$, incrementing by steps of size $0.1$.
For a recording with a tempo of 120 bpm,
a scaling of $1\pm0.1$ amounts to a change of $\pm 12$ bpm.
We then search for tempi where DeSPerF-BALLROOM
classifies all test recordings the same way.

\begin{figure}[t]
\centering
\includegraphics[height=1.9in]{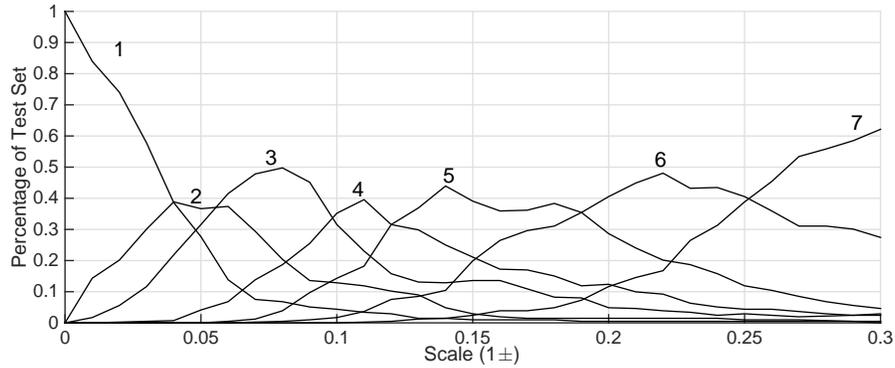}
\caption{The percentage of the {\em BALLROOM} test dataset classified by the system
in a number of different ways (numbered) as a function of the maximum scale
of frequency-preserving time stretching.
For example, with scalings in $1\pm0.08$, 
half of all test recordings are classified 3 different ways.
}
\label{fig:DeSPerF_BALLROOM_expt07_Nways}
\end{figure}

\begin{figure}[t]
\centering
\includegraphics[width=2.7in]{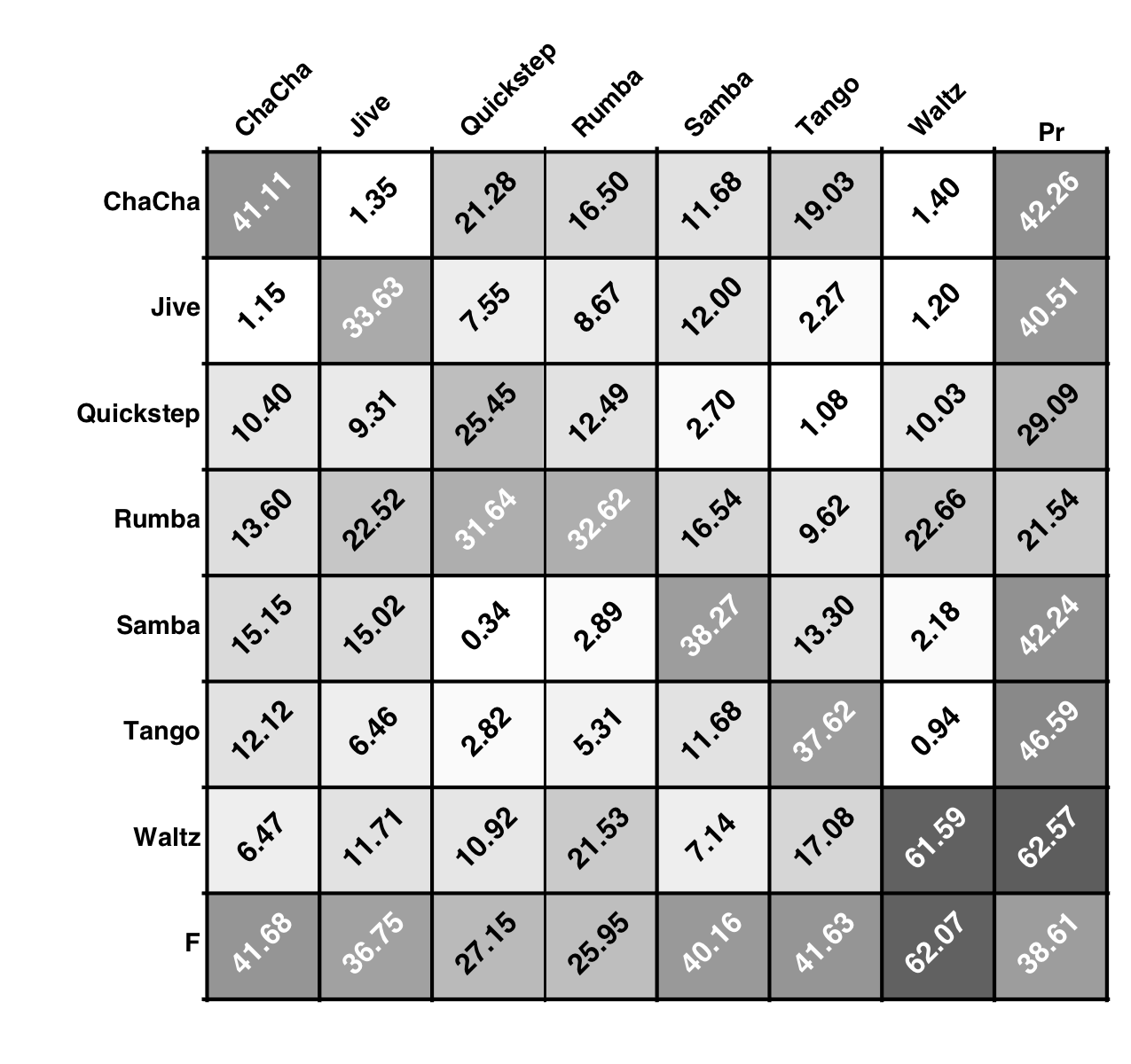}
\caption{As in Fig. \ref{fig:DeSPerF_expt00},
but for all $206$ test recordings time-stretched 
with 32 scales in $[0.85, 1.15]$.
For instance, about 47\% of all Cha cha recordings
time stretched by 32 scales in $[0.85, 1.15]$ are classified as Cha cha,
but about 6.5\% of them are classified as Waltz.}
\label{fig:DeSPerF_expt07}
\end{figure}

Figure \ref{fig:DeSPerF_BALLROOM_expt07_Nways} shows
the percentage of the test dataset classified
in a number of different ways as a function of the amount of
frequency-preserving time stretching.
With scalings between $1\pm0.1$, DeSPerF-BALLROOM
classifies about 80\% of the test dataset with 3-6 different classes.
With scalings between $1\pm0.15$, it classifies 90\% of the test recordings
into 3-7 different classes.
Figure \ref{fig:DeSPerF_expt07} shows the confusion table
of DeSPerF-BALLROOM tested with all 
$206\times 32$ time-stretched test recordings.
We see most Waltz recordings (66\%) are classified as Waltz;
however, the majority of recordings of all other labels
are classified other ways.
In the case of the Rumba recordings,
DeSPerF-BALLROOM classifies over 20\% of them as Waltz
when time stretched by at most a scale of $\pm 15$.
This entails reducing their median tempo from 100 BPM (Fig. \ref{fig:BALLROOM_tempo})
to 87, and increasing it up to 117 BPM.

We do not find tempi at which 
the system outputs the same specific class
for {\em all} test recordings.
However, we do see the following outcomes,
in order of increasing tempo:
\begin{enumerate}\small
\item DeSPerF-BALLROOM chooses Rumba for all Cha cha, Rumba, and Tango recordings
time stretched to have a tempo in the range $[95, 96.5]$ BPM;
\item DeSPerF-BALLROOM chooses Tango for all Cha cha, Jive and Tango recordings
time stretched to have a tempo in the range $[129, 130.5]$ BPM;
\item DeSPerF-BALLROOM chooses Waltz for all Cha cha and Rumba recordings 
time stretched to have a tempo in the range $[139.7, 143.7]$ BPM;
\item DeSPerF-BALLROOM chooses Samba for all Cha cha and Jive recordings
time stretched to have a tempo in the range $[144.5,147.5]$ BPM;
\item DeSPerF-BALLROOM chooses Waltz for all Cha cha and Tango recordings
time stretched to have a tempo in the range $[155.75, 157]$ BPM;
\item DeSPerF-BALLROOM chooses Cha cha for all Jive and Quickstep recordings
time stretched to have a tempo in the range $[229, 232]$ BPM.
\end{enumerate}
Clear from this is that all Cha cha
test recordings are be classified by DeSPerF-BALLROOM 
as Rumba, Samba, Tango or Waltz  
simply by changing their tempo to be in specific ranges.
This is strong evidence against the claim that 
the very high precision of DeSPerF-BALLROOM in 
Cha cha (Fig. \ref{fig:DeSPerF_expt00}) is caused by
its ability to recognise rhythmic patterns characteristic of Cha cha.

\subsection{Experiment 5: Hiring the system to compose}
The previous experiments have shown the strong
reliance of DeSPerF-BALLROOM upon cues of a temporal nature,
its inclinations toward choosing particular classes for 
recordings exhibiting different onset rates (one basic form of tempo),
the seeming class-irrelevance of larger scale stress periods (one basic form of meter),
and how it can be made to choose four other classes for any Cha cha test recording 
simply by changing only its tempo.
It is becoming more apparent that, though its FoM in Fig. \ref{fig:DeSPerF_expt00} is excellent,
we do not expect DeSPerF-BALLROOM 
to be of any use for identifying whether the music in any recording
has a particular rhythmic pattern that exists in {\em BALLROOM} --
unless one defines ``rhythmic pattern'' in a very limited way,
or claims the labels of {\em BALLROOM} are not what they seem, e.g.,
``Samba'' actually means ``any music having a tempo of 100-104 BPM.''

We now consider whether DeSPerF-BALLROOM is able to help compose
rhythmic patterns characteristic of the labels in {\em BALLROOM}.
We address this in the following way.
We randomly produce a large number of rhythmic patterns,
and synthesise recordings from them
using real audio samples of instruments typical to recordings in {\em BALLROOM}.
More specifically, for each of four voices, we generate a low-level beat structure
by sampling a Bernoulli random variable
four times for each beat in each measure (semiquaver resolution).
The parameter of the Bernoulli random variable for an onset is $p = P[1]=0.25$,
where a $1$ is an onset.
Each onset is either stressed or unstressed with equal probability.
We select a tempo sampled
from a uniform distribution over a specific range,
then synthesise repetitions of the two measures
in each voice to make a recording of 15 s.
Finally, we select as most class-representative
those recordings for which the classification of DeSPerF-BALLROOM is the most confident (\ref{eq:DNNoutput}),
and inspect how the results exemplify rhythms in {\em BALLROOM}.
This is of course a brute force approach.
We could use more sophisticated approaches
to generate compositions, such as Markov chains, e.g., \cite{Pachet2003,Thomas2013a};
but the aim of this experiment is not to produce interesting music,
but to see whether the models of DeSPerF-BALLROOM 
can confidently detect rhythmic patterns characteristic to {\em BALLROOM}.

\begin{figure}[t]
\vspace{-0.2in}
\centering 
\subfigure[Pattern 3338, classified with maximum confidence as Jive]{
\includegraphics[height=1in]{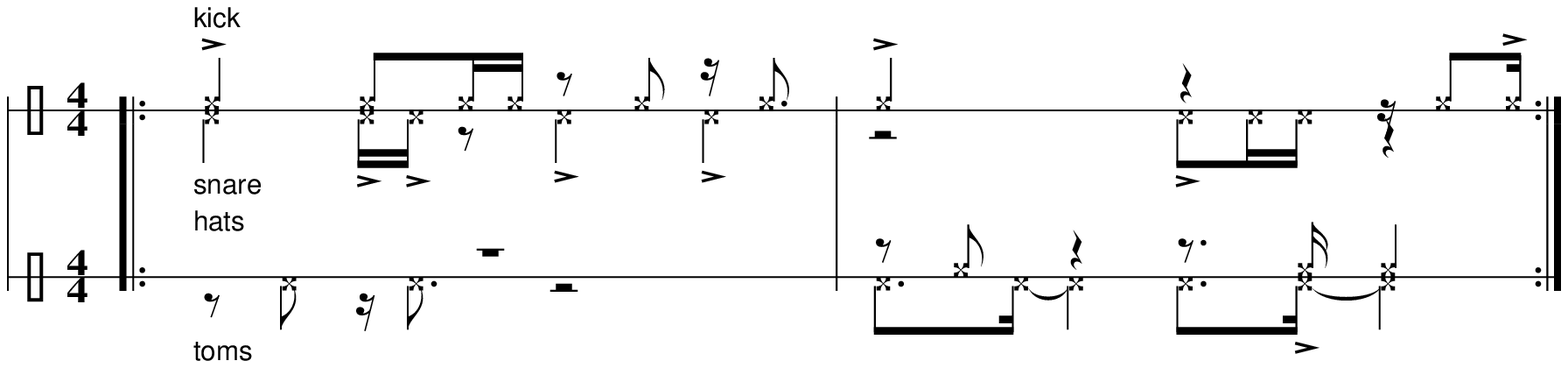}}
\subfigure[Pattern 2982, classified with maximum confidence as Quickstep]{
\includegraphics[height=1.1in]{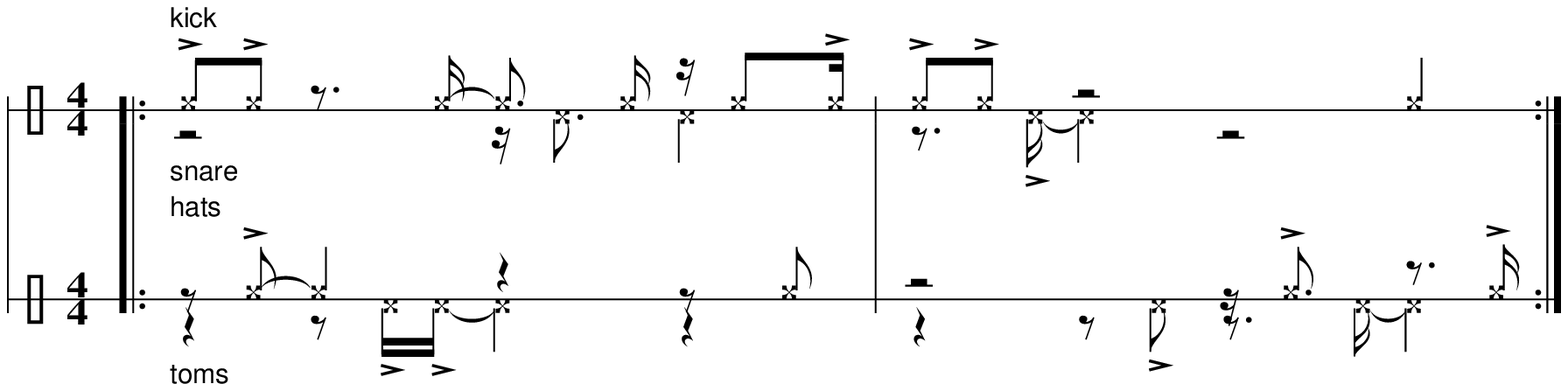}}
\subfigure[Pattern 5519, classified with maximum confidence as Tango]{
\includegraphics[height=1in]{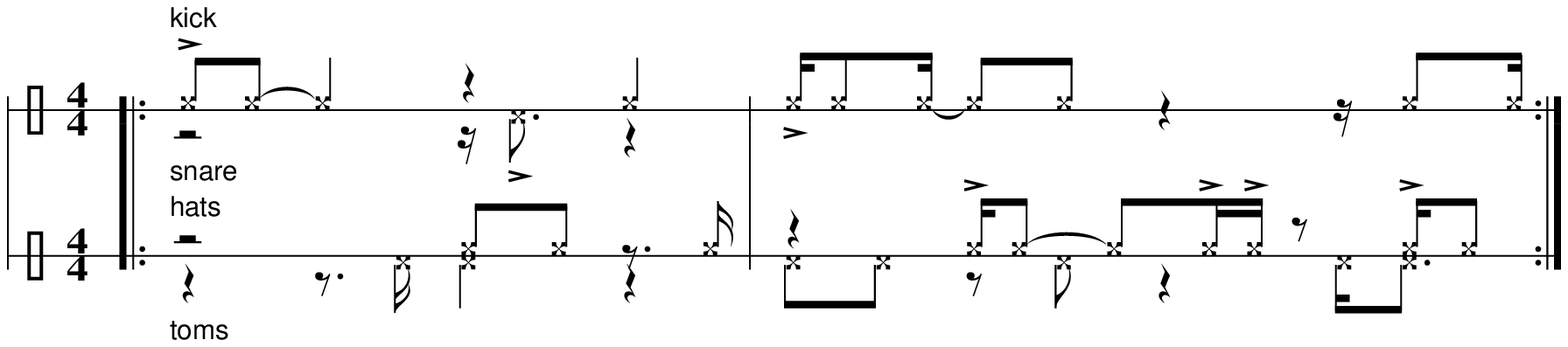}}
\subfigure[Pattern 2684, classified with maximum confidence as Waltz]{
\includegraphics[height=1.2in]{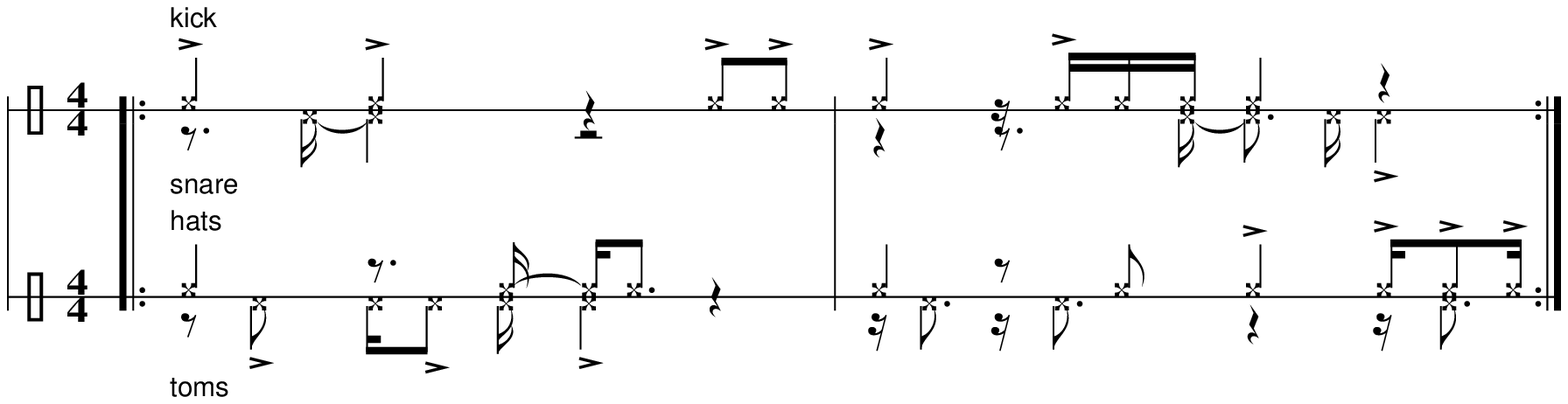}}
\caption{Some rhythmic patterns classified most confidently 
in the given class by DeSPerF-BALLROOM.}
\label{fig:patterns_synthetic}\end{figure}

To evaluate the internal model of the system for Jive,
we perform the above with audio samples of instruments typical to Jive:
kick, snare, tom, and hat.
Furthermore, we restrict the meter to be quadruple,
make sure a stressed kick occurs on 
the first beat of each measure,
and set the tempo range to $[168,176]$ BPM.
These are conditions most advantageous to 
the system, considering what it has learned about Jive
in {\em BALLROOM}.
Of 6020 synthetic recordings produced this way,
DeSPerF-BALLROOM classifies 447
with maximum confidence.
Of these, 128 are classified as Jive,
122 are classified as Waltz, 79 as Tango,
and the remainder in the four other classes.
Figure \ref{fig:patterns_synthetic} shows four of them selected at random. 
Even with these favourable settings,
it is difficult to hear in any of the recordings
similarity to the rhythmic patterns
of which they are supposedly representative.
We find similar outcomes for the other labels of {\em BALLROOM}.
In general, we find it incredibly difficult to coax anything from 
DeSPerF-BALLROOM that resembles the
rhythmic patterns in {\em BALLROOM}.

\section{Discussion}\label{sec:discussion}
To explain Fig. \ref{fig:DeSPerF_expt00},
to seek the cause of the behaviour of DeSPerF-BALLROOM,
we have dissected the system,
analysed its training and testing dataset,
and conducted several experiments.
We see from the first experiment that
the performance of DeSPerF-BALLROOM relies 
critically on cues of a temporal nature.
The results of the second experiment
reveal the inclinations of the system to 
confidently label in particular ways
recordings that all exhibit, arguably, 
the same and most simple rhythmic pattern but 
with different onset rates.
It also suggests that DeSPerF-BALLROOM is
somehow adjusting its perception of tempo,
of something highly correlated with tempo,
for recordings of some labels in {\em BALLROOM}.
The results of the third experiment
show how little the system's behaviour changes 
when we introduce longer-period repetitions
in the recordings -- a basic form of meter.
The independent variable of onset rate 
appears to trump the influence of the stress pattern.
The fourth experiment shows how
the system selects many classes for music 
exhibiting the same repetitive rhythmic patterns,
just with different tempi.
We also find some narrow tempo ranges 
in which the system classifies in the same way
all test recordings of one label.
Finally, the last experiment shows that 
the system confidently produces rhythmic patterns 
that do not clearly reflect those heard in {\em BALLROOM}.
All of this points to the conclusion that 
Fig. \ref{fig:DeSPerF_expt00} is not caused by, and does not reflect, 
an intelligence about rhythmic patterns.
The task DeSPerF-BALLROOM is performing is not
the identification of rhythmic patterns heard in music recordings.
Instead, Fig. \ref{fig:DeSPerF_expt00} appears to be caused by the 
exploitation of some cue highly related to 
the confounding of tempo with label in {\em BALLROOM},
which the system has through no fault of its own 
learned from its teaching materials.
In summary, DeSPerF-BALLROOM is
identifying rhythmic patterns as well as 
Clever Hans was solving arithmetic.

One can of course say Table \ref{tab:BALLROOMregulations}
is proof that tempo is extremely relevant 
for ballroom dance music classification.
Supported by such formal rules, as well as the increased reproduction of ground truth
observed in {\em BALLROOM} when tempo is used as a feature, \citeN{Dixon2004} write,
``tempo is one the most important features in determining dance genre''
\cite{Dixon2003,Gouyon2004}.
Hence, one is tempted to claim that though the system
uses some cue highly related to tempo, it makes little difference.
There are four problems with this claim.
First, one can argue that tempo and rhythm are intimately connected,
but in practice they seem to be treated separately.
For instance, the rhythmic pattern features proposed by \citeN{Dixon2004} 
are tempo invariant.
In their work on measuring rhythmic similarity, 
\citeN{Holzapfel2008b} use dynamic time warping 
to compare rhythms independent of tempo (further refined in \cite{Holzapfel2009}).
Second, Table \ref{tab:BALLROOMregulations} describes 
eligibility for music {\em to be allowed in a competition of particular dance styles},
and not for music or its rhythmic patterns to be given a stylistic label.
Indeed, Fig. \ref{fig:BALLROOM_tempo}
shows several recordings in {\em BALLROOM} break
the criteria set forth by \citeN{WSDF2014}.
Third, this claim moves the goal line after the fact.
Section \ref{sec:BALLROOM} shows that
though {\em BALLROOM} poses many different tasks,
the task originally intended by \citeN{Dixon2004} 
is to extract and learn ``repetitive rhythmic patterns'' 
from recorded music audio, and not to classify ballroom dance music.
Finally, the claim that tempo is extremely relevant 
for ballroom dance music classification 
works against the aims of developing music content analysis systems.
If the information or composition needs of a user involve rhythmic
patterns characteristic of ballroom dance music styles,
then DeSPerF-BALLROOM will contribute little of value
despite its impressive and human-like FoM in Fig. \ref{fig:DeSPerF_expt00}.
The hope is the DeSPerF-BALLROOM
has learned to model rhythmic patterns.
The reality is that it is not recognising rhythmic patterns.

Automatically constructing a working model, or theory,
that explains a collection of real-world music examples
has been called ``a great intellectual challenge'' 
with major repercussions \cite{Dubnov2003a}.
As observed by Eigenfeldt et al. \citeyear{Eigenfeldt2013d,Eigenfeldt2013e,Eigenfeldt2013b},
applying a machine learning algorithm to learn relationships among and rules of
the music in a dataset (corpus) 
is in the most abstract sense {\em automated meta-creation}:
a machine learns the ``rules from which to generate new art'' \cite{Eigenfeldt2014b}. 
This same sentiment is echoed in other domains,
such as computer vision \cite{Dosovitskiy2014,Nguyen2014a},
written language \cite{Shannon1998a,Ghedini2015a},
and the recent ``zero resource speech challenge,''\footnote{\url{http://www.lscp.net/persons/dupoux/bootphon/zerospeech2014/website}} 
in which a machine listening system must learn 
basic elements of spoken natural language, e.g., phonemes and words.
In fact, the automatic modelling of music style is a pursuit far older
and more successful in the symbolic domain than in the domain
of audio signal processing
\cite{Hiller1959a,Cope1991a,Roads1996,Dubnov2003a,Pachet2003,Pachet2005,Collins2010b,Argamon2010a,Dubnov2014a,Eigenfeldt2012b,Eigenfeldt2013b,Eigenfeldt2013d}.
One reason for the success of music style emulation in the symbolic domain
is that notated music is automatically  
on a plane more meaningful than samples of an audio signal,
or features derived from such basic representations. 
It is closer to ``the musical surface'' \cite{Dubnov2003a,Dubnov2014a}.
In his work on the algorithmic emulation of electronic dance music,
\citeN{Eigenfeldt2013d} highlights some severe impediments
arising from working with music audio recordings:
reliability, interpretability, and usability.
They found that
the technologies offered so far by content-based music information retrieval 
do not yet provide suitably rich and meaningful representations
from which a machine can learn about music.
\citeN{Eigenfeldt2013d} thus bypasses these problems by sacrificing scalability,
and approaching the automated style modelling of 
electronic dance music in the symbolic domain by first transcribing
by hand a corpus of dance music \cite{Eigenfeldt2013b,Eigenfeldt2013d}.

Another reason why the pursuit of style detection, understanding, and emulation
in the symbolic domain has seen substantial success 
whereas that in the audio domain has not 
is the relevance of evaluation practices in each domain.
A relevant evaluation of success toward the pursuit of music understanding
is how well a system can create ``new art'' that reflects its training \cite{Eigenfeldt2014b}.
As with the ``continuator'' \cite{Pachet2003} -- where 
a computer agent ``listens'' to the performance of a musician,
and then continues where the musician leaves off --
the one being emulated becomes the judge.
This is also the approach used by \citeN{Dannenberg1997}
in their music style recognition system,
which sidesteps the thorny issue of having to define  
what is being emulated or recognised.
Unfortunately, much research in developing music content analysis systems
has approached the evaluation of such technologies in ways that,
while convenient, widely accepted, and precise, are not relevant.
In essence, the proof of good pudding is in its eating,
not in the fact that its ingredients were precisely measured.

Among the nearly 500 publications 
about the automatic recognition of music genre or style \cite{Sturm2014d},
only a few works evaluate the internal models learned by a system 
by looking at the music it composes. 
\citeN{Cruz2003} construct a system that attempts to learn 
language models from notated music melodies in a variety of styles
(Gregorian, Baroque, Ragtime).
They implement these models as finite state automata,
and then use them to generate exemplary melodies in each style.
As in Fig. \ref{fig:patterns_synthetic}, \citeN{Cruz2003}
provide examples of the produced output,
and reflect on the quality of the results
(which they expand upon in a journal article \cite{Cruz2008}).
In the audio domain, \citeN{Sturm2012c} employs a brute force approach
to exploring the sanity of the learned models of two different 
state-of-the-art music content analysis systems
producing high FoM in a benchmark music genre dataset.
He generates random recordings from sample loops,
has each system classify them, 
and keeps only those made with high confidence.
From a listening experiment, he finds
that people cannot identify the genres of those representative excerpts.\footnote{It is 
entirely likely that I have missed relevant references
from the symbolic domain for genre/style recognition/emulation.}
In a completely different domain, similar approaches
have recently been used to test the sanity of the internal models
of high-performing image content recognition systems \cite{Szegedy2014,Dosovitskiy2014,Nguyen2014a}.

The results of our analysis and experiments with 
DeSPerF-BALLROOM clearly do not support 
rejecting the hypothesis that this system is a ``horse''
with respect to identifying rhythmic patterns;
but what about the DeSPerF-based systems that reproduced 
the most ground truth in the 2013 MIREX edition of the 
``Audio Latin Music Genre classification task'' (ALGC)?
Can we now conclude that their winning performance was not caused 
by ``musical intelligence,'' but by the exploitation of some tempo-like cue?
In the case of the {\em LMD} dataset  used in ALGC,
the task appears to be ``musical genre classification'' \cite{Silla2008b}.
\citeN{Silla2008b} reference \citeN{Fabbri1999} to define ``genre:''
``a kind of music, as it is acknowledged by a community for any reason or purpose or criteria.''
In particular to {\em LMD},
the community acknowledging these ``kinds'' of music 
was represented by two
``professional teachers with over ten years of experience 
in teaching ballroom and Brazilian cultural dances'' \cite{Silla2008b}.
These professionals selected commercial recordings of music
``that they judged representative of a specific genre, 
according to how that musical recording is danced.''
The appendix to \cite{Silla2008b} gives characteristics of 
the music genres in {\em LMD}, many of which
should be entirely outside the purview of any audio-based system,
e.g., aspects of culture, topic, geography, and dance moves.
We cannot say what the cue in {\em LMD} is --
and tempo currently does not appear to be a confound \cite{Esparza2014} --
but the default position in light of the poor evidence
contributed by the amount of ground truth reproduced 
{\em must be} that the system is not yet demonstrated
to possess the ``intelligence'' relevant for a specific task.
Valid experiments are needed to claim otherwise \cite{Urbano2013}.

The task of creating Fig. \ref{fig:patterns_BALLROOM} was laborious.
Identifying these rhythmic patterns
relies on experience in listening to mixtures of voices 
and separating instruments,
listening comparatively to collections of music recordings, 
memory, expectation, musical practice, physicality, and so on.
Constructing an artificial system that can automatically 
do something like this for an arbitrarily large collection of music audio recordings 
will surely produce major advances
in machine listening and creativity \cite{Dubnov2003a}.
In proportion, evidence for such abilities 
must be just as outstanding -- much more so
than achieving 100\% on the rather tepid multiple choice exam.
It is of course the hope that DeSPerF-BALLROOM
has learned from a collection of music recordings 
{\em general} models of the styles tersely represented by the labels;
and indeed, ``One of machine learning's main purposes is to create the capability to sensibly generalize''
\cite{Dubnov2003a}.
The results in Fig. \ref{fig:DeSPerF_expt00} 
just does not provide valid evidence for such a conclusion;
it does not even provide evidence that such capabilities are within reach.
Similarly, we are left to question all results in Table \ref{tab:BALLROOMresults}:
which of these are ``horses'' like DeSPerF-BALLROOM,
and which are solutions, for identifying rhythmic patterns?
What problem is each actually solving,
and how is it related to music?
Which can be useful for connecting users with music and 
information about music?
Which can facilitate creative pursuits?
Returning to the formalism presented in Section \ref{sec:problemofMCA},
for what {\em use cases} can each system actually benefit?
One might say that any system
using musically interpretable features is likely a solution.
For instance, the features employed by \citeN{Dixon2004}
are essentially built from bar-synchronised decimated amplitude envelopes,
and are interpretable with respect to the rhythmic characteristics of the styles in {\em BALLROOM}.
However, as seen at the end of Section \ref{sec:features},
SPerF are musically interpretable as well.
One must look under the hood,
and design, implement and analyse experiments 
that have the validity to test to the objective.

Ascribing too much importance to the measurement and comparison of 
the amounts of ground truth reproduced --
a practice that appears in a vast majority of 
publications in music genre recognition \cite{Sturm2012e,Sturm2014d} --
is an impediment to progress.
Consider a system trained and tested in {\em BALLROOM}
that has actually learned to recognise rhythmic patterns characteristic of waltz,
but has trouble with any rhythmic patterns not in triple meter.
Auditioning {\em BALLROOM}
demonstrates that all observations not labeled Waltz have a duple or quadruple meter.
If such a system correctly classifies all Waltz test recordings based on rhythmic patterns,
but chooses randomly for all others,
we expect its normalised accuracy to be about $28.5$\%.
This is double that expected of a random selection,
but far below the accuracy seen in Fig. \ref{fig:DeSPerF_expt00}.
It is thus not difficult to believe the low-performing system would be tossed
for DeSPerF-BALLROOM, or even let pass through peer review, 
even though it is actually the case that
the former system is addressing the task of rhythmic pattern recognition,
while the latter is just a ``horse.''
Such a warning has been given before: 
``an improved general music similarity algorithm might even yield lower accuracies" \cite{Pohle2009}.
{\em No system should be left behind because of invalid experiments.}

Many interesting questions arise from our work.
What will happen when SPerF are made tempo-invariant?
What will happen if the tempo confounding in {\em BALLROOM} is removed?
One can imagine augmenting the training dataset 
by performing many different pitch-preserving time stretching transformations;
or of making all recordings have the same tempo.
Will the resulting system then learn to identify 
repetitive rhythmic patterns?
Or will it only appear so by use of another cue?
Another question is what the DNN contributes?
In particular, DNN have been claimed to be able to
learn to ``listen'' to music in a hierarchical fashion 
\cite{Hamel2010,Humphrey2013,Deng2014}.
If a DNN-based system is actually addressing the task of 
identifying rhythmic patterns,
how does this hierarchical listening manifest?
Is it over beats, figures, and bars?
This also brings up the question, ``why learn at all?'' 
Should we expect the system to acquire 
what is readily available from experts?
Why not use expert knowledge, or at least
leverage automated learning with an expert-based system?
Finally, the concepts of meta-creation motivates new evaluation methods \cite{Thomas2013a},
both in determining the sanity of a system's internal models,
but also in meaningfully comparing these models.
Meta-creation essentially motivates the advice of hiring the system 
to do the accounting in order to reveal the ``horse.''
Valid evaluation approaches will undoubtedly require more effort
on the part of the music content analysis system developer,
but validity is simply non-negotiable.





%
\section{Conclusion}
The first supplement in \citeN{Pfungst1911}
describes the careful and strict methods used 
to teach the horse Clever Hans over the course of four years 
to read letters and numerals, and then to solve simple problems of arithmetic.
When Clever Hans had learned these basics, 
had time ``to discover a great deal for himself,''
and began to give solutions to unique problems
that were not part of his training, his handler believed
``he had succeeded in inculcating the inner meaning of the number concepts, 
and not merely an external association of memory images 
with certain movement responses'' \cite{Pfungst1911}.
Without knowing the story of Clever Hans,
it seems quite reasonable to conclude that 
since it is highly unlikely for DeSPerF-BALLROOM 
to achieve the FoM in Fig. \ref{fig:DeSPerF_expt00} by luck alone,
then it must have learned rhythmic patterns in 
the recorded music in {\em BALLROOM}.
As in the case of Clever Hans's tutor,
there are four problems with such a conclusion.

First, this unjustifiably anthropomorphises the system of Fig. \ref{fig:DeSPerF_expt00}.
For instance, someone who does not know better might believe that a
stereo system must be quite a capable musician
because they hear it play music.
There is no evidence that the criteria and rules used by this system --
the ones completely obfuscated by the cascade of compressed affine linear
transformations described in Section 3.2 --
are among those that a human 
uses to discriminate between and identify style in music listening.
Second, one makes the assumption that
the semantics of the labels of the dataset 
refer to some quality called ``style'' or ``rhythmic pattern.'' 
This thus equates, ``learning to map statistics of a sampled time series
to tokens,'' and ``learning to discriminate between and identify styles 
that manifest in recorded music.''
Third, underpinning this conclusion is the assumption that the tutoring 
was actually teaching the skills desired.
In the case of the system of Fig. \ref{fig:DeSPerF_expt00},
the tutoring actually proceeds by asking the DNN a question 
(inputting an element of $\UniverseSemanticFeature$ with ground truth $s \in \UniverseSemantic$),
comparing its output $\vx^{(K)}$ to the target $\ve_s$
(the standard basic vector with a $1$ in the row 
associated with $s$ and zero everywhere else),
then adapting all of its parameters in an optimal direction toward that target,
and finally repeating.
While this ``pedagogy'' is certainly strict and provably optimal 
with respect to specific objectives \cite{Deng2014,Hastie2009},
its relationship to ``learning to discriminate between and identify styles'' is not clear.
Repeatedly forcing Hans to tap his hoof twice is not so clearly
teaching him about the ``inner meaning of the number concept'' 2.
Fourth, and most significantly, this conclusion implicitly and incorrectly assumes 
that the results of Fig. \ref{fig:DeSPerF_expt00}
have only two possible explanations: luck or ``musical intelligence.''
The story of Clever Hans shows just how misguided such a belief can be.

The usefulness of any music content analysis system 
depends on what task it is actually performing,
what problem it is actually solving.
{\em BALLROOM} at first appears to explicitly pose 
a clear problem; but we now see that 
there exists several ways to reproduce its ground truth --
each of which involves a different task, e.g., rhythmic pattern recognition,
tempo detection, instrument recognition, and/or ones
that have no concrete relationship to music.
We cannot tell which task DeSPerF-BALLROOM is performing
just from looking at Fig. \ref{fig:DeSPerF_expt00}.
While comparing the output of a music content analysis system
to the ground truth of a dataset is convenient,
it simply does not distinguish between ``horses'' and solutions \cite{Sturm2012e,Sturm2013g}.
It does not produce valid evidence of intelligence.
That is, we cannot know whether 
the system is giving the right answers for the {\em wrong reasons.}
Just as Clever Hans appeared to be solving
problems of arithmetic -- what can be more explicit
than asking a horse to add 1 and 1? --
the banal task he was actually performing,
unbeknownst to many save himself, 
was ``make the nice man feed me.''
The same might be true, metaphorically speaking, 
for the systems in Table \ref{tab:BALLROOMresults}.

\begin{acks}
Thank you to Aggelos Pikrakis, Corey Kereliuk, Jan Larsen,
and the anonymous reviewers.
I dedicate this article to the memory of Alan Young (1919-2016),
principal actor of the TV show, ``Mr. Ed.''

\end{acks}


\bibliographystyle{ACM-Reference-Format-Journals}
\bibliography{../../bibliographies/genre,../../bibliographies/emotion,../../bibliographies/tagging,../../bibliographies/BibAnnon}


\begin{thebibliography}{00}


\ifx \showCODEN    \undefined \def \showCODEN     #1{\unskip}     \fi
\ifx \showDOI      \undefined \def \showDOI       #1{{\tt DOI:}\penalty0{#1}\ }
  \fi
\ifx \showISBNx    \undefined \def \showISBNx     #1{\unskip}     \fi
\ifx \showISBNxiii \undefined \def \showISBNxiii  #1{\unskip}     \fi
\ifx \showISSN     \undefined \def \showISSN      #1{\unskip}     \fi
\ifx \showLCCN     \undefined \def \showLCCN      #1{\unskip}     \fi
\ifx \shownote     \undefined \def \shownote      #1{#1}          \fi
\ifx \showarticletitle \undefined \def \showarticletitle #1{#1}   \fi
\ifx \showURL      \undefined \def \showURL       #1{#1}          \fi

\bibitem[\protect\citeauthoryear{Argamon, Burns, and Dubnov}{Argamon
  et~al\mbox{.}}{2010}]%
        {Argamon2010a}
{S. Argamon}, {K. Burns}, {and} {S. Dubnov} (Eds.). 2010.
\newblock {\em The Structure of Style: Algorithmic Approaches to Understanding
  Manner and Meaning}.
\newblock Springer.
\newblock


\bibitem[\protect\citeauthoryear{Aucouturier}{Aucouturier}{2009}]%
        {Aucouturier2009}
{J.-.J. Aucouturier}. 2009.
\newblock \showarticletitle{Sounds like teen spirit: Computational insights
  into the grounding of everyday musical terms}.
\newblock In {\em Language, Evolution and the Brain: Frontiers in Linguistic
  Series}, {J.~Minett} {and} {W.~Wang} (Eds.). Academia Sinica Press.
\newblock


\bibitem[\protect\citeauthoryear{Bergstra, Casagrande, Erhan, Eck, and
  K\'{e}gl}{Bergstra et~al\mbox{.}}{2006}]%
        {Bergstra2006}
{J. Bergstra}, {N. Casagrande}, {D. Erhan}, {D. Eck}, {and} {B. K\'{e}gl}.
  2006.
\newblock \showarticletitle{Aggregate features and {AdaBoost} for music
  classification}.
\newblock {\em Machine Learning\/} {65}, 2-3 (June 2006), 473--484.
\newblock


\bibitem[\protect\citeauthoryear{Casey, Veltkamp, Goto, Leman, Rhodes, and
  Slaney}{Casey et~al\mbox{.}}{2008}]%
        {Casey2008a}
{M. Casey}, {R. Veltkamp}, {M. Goto}, {M. Leman}, {C. Rhodes}, {and} {M.
  Slaney}. 2008.
\newblock \showarticletitle{Content-based Music Information Retrieval: Current
  Directions and Future Challenges}.
\newblock {\em Proc. IEEE\/} {96}, 4 (Apr. 2008), 668--696.
\newblock


\bibitem[\protect\citeauthoryear{Collins}{Collins}{2010}]%
        {Collins2010b}
{Nick Collins}. 2010.
\newblock {\em Introduction to Computer Music}.
\newblock Wiley.
\newblock


\bibitem[\protect\citeauthoryear{Cope}{Cope}{1991}]%
        {Cope1991a}
{D. Cope}. 1991.
\newblock {\em Computers and Musical Style}.
\newblock Oxford University Press.
\newblock


\bibitem[\protect\citeauthoryear{Cruz and Vidal-Ruiz}{Cruz and
  Vidal-Ruiz}{2003}]%
        {Cruz2003}
{P.P. Cruz} {and} {E. Vidal-Ruiz}. 2003.
\newblock \showarticletitle{Modeling musical style using grammatical inference
  techniques: a tool for classifying and generating melodies}. In {\em Proc.
  WEDELMUSIC}. 77--84.
\newblock
\showDOI{%
\url{http://dx.doi.org/10.1109/WDM.2003.1233878}}


\bibitem[\protect\citeauthoryear{Cruz and Vidal}{Cruz and Vidal}{2008}]%
        {Cruz2008}
{Pedro~P. Cruz} {and} {Enrique Vidal}. 2008.
\newblock \showarticletitle{TWO GRAMMATICAL INFERENCE APPLICATIONS IN MUSIC
  PROCESSING.}
\newblock {\em Applied Artificial Intell.\/} {22}, 1/2 (2008), 53--76.
\newblock


\bibitem[\protect\citeauthoryear{Cunningham, Bainbridge, and Downie}{Cunningham
  et~al\mbox{.}}{2012}]%
        {Cunningham2012}
{S.~J. Cunningham}, {D. Bainbridge}, {and} {J.~S. Downie}. 2012.
\newblock \showarticletitle{The Impact of {MIREX} on Scholarly Research}. In
  {\em Proc. ISMIR}. 259--264.
\newblock


\bibitem[\protect\citeauthoryear{Dannenberg, Thom, and Watson}{Dannenberg
  et~al\mbox{.}}{1997}]%
        {Dannenberg1997}
{R.~B. Dannenberg}, {B. Thom}, {and} {D. Watson}. 1997.
\newblock \showarticletitle{A Machine Learning Approach to Musical Style
  Recognition}. In {\em Proc. ICMC}. 344--347.
\newblock


\bibitem[\protect\citeauthoryear{Deng and Yu}{Deng and Yu}{2014}]%
        {Deng2014}
{L. Deng} {and} {D. Yu}. 2014.
\newblock {\em Deep Learning: Methods and Applications}.
\newblock Now Publishers.
\newblock


\bibitem[\protect\citeauthoryear{Dixon, Gouyon, and Widmer}{Dixon
  et~al\mbox{.}}{2004}]%
        {Dixon2004}
{S. Dixon}, {F. Gouyon}, {and} {G. Widmer}. 2004.
\newblock \showarticletitle{Towards characterisation of music via rhythmic
  patterns}. In {\em Proc. ISMIR}. 509--517.
\newblock


\bibitem[\protect\citeauthoryear{Dixon, Pampalk, and Widmer}{Dixon
  et~al\mbox{.}}{2003}]%
        {Dixon2003}
{S. Dixon}, {E. Pampalk}, {and} {G. Widmer}. 2003.
\newblock \showarticletitle{Classification of Dance Music by Periodicity
  Patterns}. In {\em Proc. ISMIR}.
\newblock


\bibitem[\protect\citeauthoryear{Dosovitskiy, Springenberg, and
  Brox}{Dosovitskiy et~al\mbox{.}}{2014}]%
        {Dosovitskiy2014}
{Alexey Dosovitskiy}, {Jost~Tobias Springenberg}, {and} {Thomas Brox}. 2014.
\newblock \showarticletitle{Learning to Generate Chairs with Convolutional
  Neural Networks}.
\newblock {\em CoRR\/}  {abs/1411.5928} (2014).
\newblock


\bibitem[\protect\citeauthoryear{Downie, Ehmann, Bay, and Jones}{Downie
  et~al\mbox{.}}{2010}]%
        {Downie2010}
{J. Downie}, {Andreas Ehmann}, {Mert Bay}, {and} {M. Jones}. 2010.
\newblock \showarticletitle{The Music Information Retrieval Evaluation
  eXchange: Some Observations and Insights}.
\newblock In {\em Advances in Music Information Retrieval}, {Zbigniew Ras}
  {and} {Alicja Wieczorkowska} (Eds.). Springer Berlin / Heidelberg, 93--115.
\newblock


\bibitem[\protect\citeauthoryear{Downie}{Downie}{2004}]%
        {Downie2004b}
{J.~S. Downie}. 2004.
\newblock \showarticletitle{The scientific evaluation of music information
  retrieval systems: Foundations and future}.
\newblock {\em Computer Music Journal\/} {28}, 2 (2004), 12--23.
\newblock


\bibitem[\protect\citeauthoryear{Downie}{Downie}{2008}]%
        {Downie2008}
{J.~S. Downie}. 2008.
\newblock \showarticletitle{The music information retrieval evaluation exchange
  (2005--2007): A window into music information retrieval research}.
\newblock {\em Acoustical Science and Tech.\/} {29}, 4 (2008), 247--255.
\newblock


\bibitem[\protect\citeauthoryear{Dubnov, Assayag, Lartillot, and
  Bejerano}{Dubnov et~al\mbox{.}}{2003}]%
        {Dubnov2003a}
{S. Dubnov}, {G. Assayag}, {O. Lartillot}, {and} {G. Bejerano}. 2003.
\newblock \showarticletitle{Using machine-learning methods for musical style
  modeling}.
\newblock {\em Computer\/} {36}, 10 (2003), 73--80.
\newblock


\bibitem[\protect\citeauthoryear{Dubnov and Surges}{Dubnov and Surges}{2014}]%
        {Dubnov2014a}
{S. Dubnov} {and} {G. Surges}. 2014.
\newblock {\em Digital Da Vinci}.
\newblock Springer, Chapter Delegating Creativity: Use of Musical Algorithms in
  Machine Listening and Composition, 127--158.
\newblock


\bibitem[\protect\citeauthoryear{Eigenfeldt}{Eigenfeldt}{2012}]%
        {Eigenfeldt2012b}
{A. Eigenfeldt}. 2012.
\newblock \showarticletitle{Embracing the Bias of the Machine: Exploring
  Non-Human Fitness Functions}. In {\em Proceedings of the AAAI Conference on
  Artificial Intelligence and Interactive Digital Entertainment}.
\newblock


\bibitem[\protect\citeauthoryear{Eigenfeldt}{Eigenfeldt}{2013}]%
        {Eigenfeldt2013d}
{A. Eigenfeldt}. 2013.
\newblock \showarticletitle{The Human Fingerprint in Machine Generated Music}.
  In {\em Proc. xCoAx}.
\newblock


\bibitem[\protect\citeauthoryear{Eigenfeldt and Pasquier}{Eigenfeldt and
  Pasquier}{2013a}]%
        {Eigenfeldt2013e}
{A. Eigenfeldt} {and} {P. Pasquier}. 2013a.
\newblock \showarticletitle{Considering vertical and horizontal context in
  corpus-based generative electronic dance music}. In {\em Proc. Int. Conf.
  Computational Creativity}.
\newblock


\bibitem[\protect\citeauthoryear{Eigenfeldt and Pasquier}{Eigenfeldt and
  Pasquier}{2013b}]%
        {Eigenfeldt2013b}
{A. Eigenfeldt} {and} {P. Pasquier}. 2013b.
\newblock \showarticletitle{Evolving Structures for Electronic Dance Music}. In
  {\em Proc. Conf. Genetic and Evolutionary Computation}. 319--326.
\newblock


\bibitem[\protect\citeauthoryear{Eigenfeldt, Thorogood, Bizzocchi, Pasquier,
  and Calvert}{Eigenfeldt et~al\mbox{.}}{2014}]%
        {Eigenfeldt2014b}
{A. Eigenfeldt}, {M. Thorogood}, {J. Bizzocchi}, {P. Pasquier}, {and} {T.
  Calvert}. 2014.
\newblock \showarticletitle{Video, Music and Sound Metacreation}. In {\em Proc.
  xCoAx}. Porto, Portugal.
\newblock


\bibitem[\protect\citeauthoryear{Esparza, Bello, and Humphrey}{Esparza
  et~al\mbox{.}}{2014}]%
        {Esparza2014}
{T.~M. Esparza}, {J.~P. Bello}, {and} {E.~J. Humphrey}. 2014.
\newblock \showarticletitle{From genre classification to rhythm similarity:
  Computational and musicological insights}.
\newblock {\em J. New Music Research\/} (2014).
\newblock


\bibitem[\protect\citeauthoryear{Fabbri}{Fabbri}{1999}]%
        {Fabbri1999}
{F. Fabbri}. 1999.
\newblock \showarticletitle{Browsing Musical Spaces: Categories and the musical
  mind}. In {\em Proc. Int. Association for the Study of Popular Music}.
\newblock


\bibitem[\protect\citeauthoryear{Flexer, Gouyon, Dixon, and Widmer}{Flexer
  et~al\mbox{.}}{2006}]%
        {Flexer2006}
{A. Flexer}, {F. Gouyon}, {S. Dixon}, {and} {G. Widmer}. 2006.
\newblock \showarticletitle{Probabilistic Combination of Features for Music
  Classification}. In {\em Proc. ISMIR}. Victoria, BC, Canda, 111--114.
\newblock


\bibitem[\protect\citeauthoryear{Fu, Lu, Ting, and Zhang}{Fu
  et~al\mbox{.}}{2011}]%
        {Fu2011}
{Z. Fu}, {G. Lu}, {K.~M. Ting}, {and} {D. Zhang}. 2011.
\newblock \showarticletitle{A Survey of Audio-Based Music Classification and
  Annotation}.
\newblock {\em IEEE Trans. Multimedia\/} {13}, 2 (Apr. 2011), 303--319.
\newblock


\bibitem[\protect\citeauthoryear{Ghedini, Pachet, and Roy}{Ghedini
  et~al\mbox{.}}{2015}]%
        {Ghedini2015a}
{F. Ghedini}, {F. Pachet}, {and} {P. Roy}. 2015.
\newblock {\em Multidisciplinary Contributions to the Science of Creative
  Thinking}.
\newblock Springer, Chapter Creating Music and Texts with Flow Machines.
\newblock


\bibitem[\protect\citeauthoryear{Goehr}{Goehr}{1994}]%
        {Goehr1994}
{L. Goehr}. 1994.
\newblock {\em The Imaginary Museum of Musical Works: An Essay in the
  Philosophy of Music}.
\newblock Oxford University Press.
\newblock


\bibitem[\protect\citeauthoryear{Gouyon}{Gouyon}{2005}]%
        {Gouyon2005}
{F. Gouyon}. 2005.
\newblock {\em A computational approach to rhythm description --- Audio
  features for the computation of rhythm periodicity functions and their use in
  tempo induction and music content processing}.
\newblock Ph.D. Dissertation. Universitat Pompeu Fabra.
\newblock


\bibitem[\protect\citeauthoryear{Gouyon and Dixon}{Gouyon and Dixon}{2004}]%
        {Gouyon2004b}
{F. Gouyon} {and} {S. Dixon}. 2004.
\newblock \showarticletitle{Dance Music Classification: A Tempo-Based
  Approach}. In {\em Proc. ISMIR}. 501--504.
\newblock


\bibitem[\protect\citeauthoryear{Gouyon, Dixon, Pampalk, and Widmer}{Gouyon
  et~al\mbox{.}}{2004}]%
        {Gouyon2004}
{F. Gouyon}, {S. Dixon}, {E. Pampalk}, {and} {G. Widmer}. 2004.
\newblock \showarticletitle{Evaluating rhythmic descriptors for musical genre
  classification}. In {\em Proc. Int. Audio Eng. Soc. Conf.} 196--204.
\newblock


\bibitem[\protect\citeauthoryear{Gouyon, Sturm, Oliveira, Hespanhol, and
  Langlois}{Gouyon et~al\mbox{.}}{2013}]%
        {Gouyon2013}
{F. Gouyon}, {B.~L. Sturm}, {J.~L. Oliveira}, {N. Hespanhol}, {and} {T.
  Langlois}. 2013.
\newblock \showarticletitle{On evaluation in music autotagging research}.
\newblock {\em (submitted)\/} (2013).
\newblock


\bibitem[\protect\citeauthoryear{Hamel and Eck}{Hamel and Eck}{2010}]%
        {Hamel2010}
{P. Hamel} {and} {D. Eck}. 2010.
\newblock \showarticletitle{Learning features from music audio with deep belief
  networks}. In {\em Proc. ISMIR}. 339--344.
\newblock


\bibitem[\protect\citeauthoryear{Hastie, Tibshirani, and Friedman}{Hastie
  et~al\mbox{.}}{2009}]%
        {Hastie2009}
{T. Hastie}, {R. Tibshirani}, {and} {J. Friedman}. 2009.
\newblock {\em The Elements of Statistical Learning: Data Mining, Inference,
  and Prediction\/} (2 ed.).
\newblock Springer-Verlag.
\newblock


\bibitem[\protect\citeauthoryear{Hiller and Isaacson}{Hiller and
  Isaacson}{1959}]%
        {Hiller1959a}
{L. Hiller} {and} {L. Isaacson}. 1959.
\newblock {\em Experimental Music: Composition with an Electronic Computer}.
\newblock Greenwood Press.
\newblock


\bibitem[\protect\citeauthoryear{Holzapfel and Stylianou}{Holzapfel and
  Stylianou}{2008}]%
        {Holzapfel2008b}
{A. Holzapfel} {and} {Y. Stylianou}. 2008.
\newblock \showarticletitle{Rhythmic similarity of music based on dynamic
  periodicity warping}. In {\em Proc. ICASSP}. 2217--2220.
\newblock


\bibitem[\protect\citeauthoryear{Holzapfel and Stylianou}{Holzapfel and
  Stylianou}{2009}]%
        {Holzapfel2009}
{A. Holzapfel} {and} {Y. Stylianou}. 2009.
\newblock \showarticletitle{A Scale Based Method for Rhythmic Similarity of
  Music}. In {\em Proc. ICASSP}. Taipei, Taiwan, 317--320.
\newblock


\bibitem[\protect\citeauthoryear{Humphrey, Bello, and LeCun}{Humphrey
  et~al\mbox{.}}{2013}]%
        {Humphrey2013}
{E.~J. Humphrey}, {J.~P. Bello}, {and} {Y. LeCun}. 2013.
\newblock \showarticletitle{Feature Learning and Deep Architectures: New
  Directions for Music Informatics}.
\newblock {\em J. Intell. Info. Systems\/} {41}, 3 (2013), 461--481.
\newblock


\bibitem[\protect\citeauthoryear{Krebs, B{\"o}ch, and Widmer}{Krebs
  et~al\mbox{.}}{2013}]%
        {Krebs2013}
{F. Krebs}, {S. B{\"o}ch}, {and} {G. Widmer}. 2013.
\newblock \showarticletitle{Rhythmic Pattern Modeling for Beat and Downbeat
  Tracking in Musical Audio}. In {\em Proc. ISMIR}.
\newblock


\bibitem[\protect\citeauthoryear{Lidy}{Lidy}{2006}]%
        {Lidy2006}
{T. Lidy}. 2006.
\newblock {\em Evaluation of New Audio Features and Their Utilization in Novel
  Music Retrieval Applications}.
\newblock Master's\ thesis. Vienna University of Tech., Vienna, Austria.
\newblock


\bibitem[\protect\citeauthoryear{Lidy, Mayer, Rauber, de~Leon, Pertusa, and
  Quereda}{Lidy et~al\mbox{.}}{2010}]%
        {Lidy2010b}
{T. Lidy}, {R. Mayer}, {A. Rauber}, {P.~P. de Leon}, {A. Pertusa}, {and} {J.
  Quereda}. 2010.
\newblock \showarticletitle{A Cartesian Ensemble of Feature Subspace
  Classifiers for Music Categorization}. In {\em Proc. ISMIR}. 279--284.
\newblock


\bibitem[\protect\citeauthoryear{Lidy and Rauber}{Lidy and Rauber}{2005}]%
        {Lidy2005}
{T. Lidy} {and} {A. Rauber}. 2005.
\newblock \showarticletitle{Evaluation of Feature Extractors and
  Psycho-Acoustic Transformations for Music Genre Classification}. In {\em
  Proc. ISMIR}. 34--41.
\newblock


\bibitem[\protect\citeauthoryear{Lidy and Rauber}{Lidy and Rauber}{2008}]%
        {Lidy2008}
{Thomas Lidy} {and} {Andreas Rauber}. 2008.
\newblock \showarticletitle{Classification and Clustering of Music for Novel
  Music Access Applications}.
\newblock In {\em Machine Learning Techniques for Multimedia}, {Matthieu Cord}
  {and} {P{\'a}draig Cunningham} (Eds.). Springer Berlin / Heidelberg,
  249--285.
\newblock
\showISBNx{978-3-540-75171-7}


\bibitem[\protect\citeauthoryear{Lidy, Rauber, Pertusa, and nesta}{Lidy
  et~al\mbox{.}}{2007}]%
        {Lidy2007}
{T. Lidy}, {A. Rauber}, {A. Pertusa}, {and} {J.~M.~I\ nesta}. 2007.
\newblock \showarticletitle{Improving Genre Classification by Combination of
  Audio and Symbolic Descriptors Using a Transcription System}. In {\em Proc.
  ISMIR}. Vienna, Austria, 61--66.
\newblock
\showISBNx{978-3-85403-218-2}


\bibitem[\protect\citeauthoryear{Mayer, Rauber, Ponce~de Le\'{o}n,
  P{\'e}rez-Sancho, and I\~{n}esta}{Mayer et~al\mbox{.}}{2010}]%
        {Mayer2010c}
{R. Mayer}, {A. Rauber}, {P.~J. Ponce~de Le\'{o}n}, {C. P{\'e}rez-Sancho},
  {and} {J.~M. I\~{n}esta}. 2010.
\newblock \showarticletitle{Feature selection in a cartesian ensemble of
  feature subspace classifiers for music categorisation}. In {\em Proc. ACM
  Int. Workshop Machine Learning and Music}. 53--56.
\newblock
\showISBNx{978-1-4503-0161-9}
\showDOI{%
\url{http://dx.doi.org/10.1145/1878003.1878021}}


\bibitem[\protect\citeauthoryear{Nguyen, Yosinski, and Clune}{Nguyen
  et~al\mbox{.}}{2015}]%
        {Nguyen2014a}
{A. Nguyen}, {J. Yosinski}, {and} {J. Clune}. 2015.
\newblock \showarticletitle{Deep neural networks are easily fooled: High
  confidence predictions for unrecognizable images}. In {\em Proc. CVPR}.
  427--436.
\newblock


\bibitem[\protect\citeauthoryear{Pachet}{Pachet}{2003}]%
        {Pachet2003}
{Francois Pachet}. 2003.
\newblock \showarticletitle{The Continuator: Musical Interaction with Style}.
\newblock {\em Journal of New Music Research\/} {32}, 3 (2003), 333--341.
\newblock


\bibitem[\protect\citeauthoryear{Pachet}{Pachet}{2011}]%
        {Pachet2005}
{F. Pachet}. 2011.
\newblock \showarticletitle{Musical Metadata and Knowledge Management}.
\newblock In {\em Encyclopedia of Knowledge Management}, {David~G. Schwartz}
  {and} {Dov Te'eni} (Eds.). IGI Global, 1192--1199.
\newblock
\showISBNx{9781599049311}
\showURL{%
\url{http://citeseerx.ist.psu.edu/viewdoc/download?doi=10.1.1.75.6804&rep=rep1&type=pdf}}


\bibitem[\protect\citeauthoryear{Peeters}{Peeters}{2005}]%
        {Peeters2005}
{G. Peeters}. 2005.
\newblock \showarticletitle{Rhythm classification using spectral rhythm
  patterns}. In {\em Proc. ISMIR}.
\newblock


\bibitem[\protect\citeauthoryear{Peeters}{Peeters}{2011}]%
        {Peeters2011}
{G. Peeters}. 2011.
\newblock \showarticletitle{Spectral and Temporal Periodicity Representations
  of Rhythm for the Automatic Classification of Music Audio Signal}.
\newblock {\em IEEE Trans. Audio, Speech, Lang. Process.\/} {19}, 5 (July
  2011), 1242--1252.
\newblock


\bibitem[\protect\citeauthoryear{Pfungst}{Pfungst}{1911}]%
        {Pfungst1911}
{O. Pfungst}. 1911.
\newblock {\em Clever Hans (The horse of Mr. Von Osten): A contribution to
  experimental animal and human psychology}.
\newblock Henry Holt, New York.
\newblock


\bibitem[\protect\citeauthoryear{Pikrakis}{Pikrakis}{2013}]%
        {Pikrakis2013}
{A. Pikrakis}. 2013.
\newblock \showarticletitle{A deep learning approach to rhythm modeling with
  applications}. In {\em Proc. Int. Workshop Machine Learning and Music}.
\newblock


\bibitem[\protect\citeauthoryear{Pohle, Schnitzer, Schedl, Knees, and
  Widmer}{Pohle et~al\mbox{.}}{2009}]%
        {Pohle2009}
{T. Pohle}, {D. Schnitzer}, {M. Schedl}, {P. Knees}, {and} {G. Widmer}. 2009.
\newblock \showarticletitle{On rhythm and general music similarity}. In {\em
  Proc. ISMIR}.
\newblock


\bibitem[\protect\citeauthoryear{Roads}{Roads}{1996}]%
        {Roads1996}
{C. Roads}. 1996.
\newblock {\em Computer Music Tutorial}.
\newblock The MIT Press.
\newblock


\bibitem[\protect\citeauthoryear{Scaringella, Zoia, and Mlynek}{Scaringella
  et~al\mbox{.}}{2006}]%
        {Scaringella2006}
{N. Scaringella}, {G. Zoia}, {and} {D. Mlynek}. 2006.
\newblock \showarticletitle{Automatic Genre Classification of Music Content: A
  Survey}.
\newblock {\em IEEE Signal Process. Mag.\/} {23}, 2 (Mar. 2006), 133--141.
\newblock


\bibitem[\protect\citeauthoryear{Schindler and Rauber}{Schindler and
  Rauber}{2012}]%
        {Schindler2012b}
{A. Schindler} {and} {A. Rauber}. 2012.
\newblock \showarticletitle{Capturing the temporal domain in Echonest Features
  for improved classification effectiveness}. In {\em Proc. Adaptive Multimedia
  Retrieval}.
\newblock


\bibitem[\protect\citeauthoryear{Schl\"uter and Osendorfer}{Schl\"uter and
  Osendorfer}{2011}]%
        {Schluter2011}
{J. Schl\"uter} {and} {C. Osendorfer}. 2011.
\newblock \showarticletitle{Music Similarity Estimation with the
  Mean-Covariance Restricted Boltzmann Machine}. In {\em Proc. ICMLA}.
\newblock


\bibitem[\protect\citeauthoryear{Schnitzer, Flexer, Schedl, and
  Widmer}{Schnitzer et~al\mbox{.}}{2011}]%
        {Schnitzer2011}
{Dominik Schnitzer}, {Arthur Flexer}, {Markus Schedl}, {and} {Gerhard Widmer}.
  2011.
\newblock \showarticletitle{Using Mutual Proximity to Improve Content-Based
  Audio Similarity}. In {\em ISMIR}. 79--84.
\newblock


\bibitem[\protect\citeauthoryear{Schnitzer, Flexer, Schedl, and
  Widmer}{Schnitzer et~al\mbox{.}}{2012}]%
        {Schnitzer2012}
{D. Schnitzer}, {A. Flexer}, {M. Schedl}, {and} {G. Widmer}. 2012.
\newblock \showarticletitle{Local and global scaling reduce hubs in space}.
\newblock {\em J. Machine Learning Res.\/}  {13} (2012), 2871--2902.
\newblock


\bibitem[\protect\citeauthoryear{Seyerlehner}{Seyerlehner}{2010}]%
        {Seyerlehner2010}
{K. Seyerlehner}. 2010.
\newblock {\em Content-based Music Recommender Systems: Beyond Simple
  Frame-level Audio Similarity}.
\newblock Ph.D. Dissertation. Johannes Kepler University, Linz, Austria.
\newblock


\bibitem[\protect\citeauthoryear{Seyerlehner, Schedl, Sonnleitner, Hauger, and
  Ionescu}{Seyerlehner et~al\mbox{.}}{2012}]%
        {Seyerlehner2012}
{K. Seyerlehner}, {M. Schedl}, {R. Sonnleitner}, {D. Hauger}, {and} {B.
  Ionescu}. 2012.
\newblock \showarticletitle{From improved auto-taggers to improved music
  similarity measures}. In {\em Proc. Adaptive Multimedia Retrieval}.
  Copenhagen, Denmark.
\newblock


\bibitem[\protect\citeauthoryear{Seyerlehner, Widmer, and Pohle}{Seyerlehner
  et~al\mbox{.}}{2010}]%
        {Seyerlehner2010b}
{K. Seyerlehner}, {G. Widmer}, {and} {T. Pohle}. 2010.
\newblock \showarticletitle{Fusing block-level features for music similarity
  estimation}. In {\em Proc. DAFx}. 1--8.
\newblock


\bibitem[\protect\citeauthoryear{Shannon and Weaver}{Shannon and
  Weaver}{1998}]%
        {Shannon1998a}
{C.~E. Shannon} {and} {W. Weaver}. 1998.
\newblock {\em The Mathematical Theory of Communication}.
\newblock University of Illinois Press.
\newblock


\bibitem[\protect\citeauthoryear{Silla, Koerich, and Kaestner}{Silla
  et~al\mbox{.}}{2008}]%
        {Silla2008b}
{C.~N. Silla}, {A.~L. Koerich}, {and} {C.~A.~A. Kaestner}. 2008.
\newblock \showarticletitle{The {Latin} Music Database}. In {\em Proc. ISMIR}.
  451--456.
\newblock


\bibitem[\protect\citeauthoryear{Slaney}{Slaney}{1998}]%
        {Slaney1998}
{M. Slaney}. 1998.
\newblock {\em Auditory Toolbox}.
\newblock {T}echnical {R}eport. Interval Research Corporation.
\newblock


\bibitem[\protect\citeauthoryear{Song, Dixon, and Pearce}{Song
  et~al\mbox{.}}{2012}]%
        {Song2012}
{Y. Song}, {S. Dixon}, {and} {M. Pearce}. 2012.
\newblock \showarticletitle{Evaluation of Musical Features for Emotion
  Classification}. In {\em Proc. ISMIR}.
\newblock


\bibitem[\protect\citeauthoryear{Sturm}{Sturm}{2012}]%
        {Sturm2012c}
{B.~L. Sturm}. 2012.
\newblock \showarticletitle{Two Systems for Automatic Music Genre Recognition:
  What Are They Really Recognizing?}. In {\em Proc. ACM MIRUM Workshop}.
  69--74.
\newblock


\bibitem[\protect\citeauthoryear{Sturm}{Sturm}{2013}]%
        {Sturm2012e}
{B.~L. Sturm}. 2013.
\newblock \showarticletitle{Classification Accuracy Is Not Enough: On the
  Evaluation of Music Genre Recognition Systems}.
\newblock {\em J. Intell. Info. Systems\/} {41}, 3 (2013), 371--406.
\newblock


\bibitem[\protect\citeauthoryear{Sturm}{Sturm}{2014a}]%
        {Sturm2013g}
{B.~L. Sturm}. 2014a.
\newblock \showarticletitle{A simple method to determine if a music information
  retrieval system is a ``horse''}.
\newblock {\em IEEE Trans. Multimedia\/} {16}, 6 (2014), 1636--1644.
\newblock


\bibitem[\protect\citeauthoryear{Sturm}{Sturm}{2014b}]%
        {Sturm2013h}
{B.~L. Sturm}. 2014b.
\newblock \showarticletitle{The State of the Art Ten Years After a State of the
  Art: Future Research in Music Information Retrieval}.
\newblock {\em J. New Music Research\/} {43}, 2 (2014), 147--172.
\newblock


\bibitem[\protect\citeauthoryear{Sturm}{Sturm}{2014c}]%
        {Sturm2014d}
{B.~L. Sturm}. 2014c.
\newblock \showarticletitle{A Survey of Evaluation in Music Genre Recognition}.
  In {\em Adaptive Multimedia Retrieval: Semantics, Context, and Adaptation},
  {A.~N\"urnberger}, {S.~Stober}, {B.~Larsen}, {and} {M.~Detyniecki} (Eds.),
  Vol. LNCS 8382. 29--66.
\newblock


\bibitem[\protect\citeauthoryear{Sturm, Kereliuk, and Pikrakis}{Sturm
  et~al\mbox{.}}{2014}]%
        {Sturm2014}
{B.~L. Sturm}, {C. Kereliuk}, {and} {A. Pikrakis}. 2014.
\newblock \showarticletitle{A Closer Look at Deep Learning Neural Networks with
  Low-level Spectral Periodicity Features}. In {\em Proc. Int. Workshop on
  Cognitive Info. Process.} 1--6.
\newblock


\bibitem[\protect\citeauthoryear{Su, Yeh, Liu, Wang, and Yang}{Su
  et~al\mbox{.}}{2014}]%
        {Su2014}
{Li Su}, {C.-C.M. Yeh}, {Jen-Yu Liu}, {Ju-Chiang Wang}, {and} {Yi-Hsuan Yang}.
  2014.
\newblock \showarticletitle{A Systematic Evaluation of the Bag-of-Frames
  Representation for Music Information Retrieval}.
\newblock {\em Multimedia, IEEE Transactions on\/} {16}, 5 (Aug 2014),
  1188--1200.
\newblock
\showISSN{1520-9210}
\showDOI{%
\url{http://dx.doi.org/10.1109/TMM.2014.2311016}}


\bibitem[\protect\citeauthoryear{Szegedy, Zaremba, Sutskever, Bruna, Erhan,
  Goodfellow, and Fergus}{Szegedy et~al\mbox{.}}{2014}]%
        {Szegedy2014}
{C. Szegedy}, {W. Zaremba}, {I. Sutskever}, {J. Bruna}, {D. Erhan}, {I.
  Goodfellow}, {and} {R. Fergus}. 2014.
\newblock \showarticletitle{Intriguing properties of neural networks}. In {\em
  Proc. ICLR}.
\newblock


\bibitem[\protect\citeauthoryear{Thomas, Pasquier, Eigenfeldt, and
  Maxwell}{Thomas et~al\mbox{.}}{2013}]%
        {Thomas2013a}
{N.~G. Thomas}, {P. Pasquier}, {A. Eigenfeldt}, {and} {J.~B. Maxwell}. 2013.
\newblock \showarticletitle{A methodology for the comparuison of melodic
  generation models using meta-melo}. In {\em Proc. ISMIR}.
\newblock


\bibitem[\protect\citeauthoryear{Tsunoo, Tzanetakis, Ono, and Sagayama}{Tsunoo
  et~al\mbox{.}}{2009}]%
        {Tsunoo2009b}
{E. Tsunoo}, {G. Tzanetakis}, {N. Ono}, {and} {S. Sagayama}. 2009.
\newblock \showarticletitle{Audio Genre Classification Using Percussive Pattern
  Clustering Combined with Timbral Features}. In {\em Proc. ICME}.
\newblock


\bibitem[\protect\citeauthoryear{Tsunoo, Tzanetakis, Ono, and Sagayama}{Tsunoo
  et~al\mbox{.}}{2011}]%
        {Tsunoo2011}
{E. Tsunoo}, {G. Tzanetakis}, {N. Ono}, {and} {S. Sagayama}. 2011.
\newblock \showarticletitle{Beyond Timbral Statistics: Improving Music
  Classification Using Percussive Patterns and Bass Lines}.
\newblock {\em IEEE Trans. Audio, Speech, and Lang. Process.\/} {19}, 4 (May
  2011), 1003--1014.
\newblock


\bibitem[\protect\citeauthoryear{Turnbull, Barrington, Torres, and
  Lanckriet}{Turnbull et~al\mbox{.}}{2008}]%
        {Turnbull2008}
{D. Turnbull}, {L. Barrington}, {D. Torres}, {and} {G. Lanckriet}. 2008.
\newblock \showarticletitle{Semantic annotation and retrieval of music and
  sound effects}.
\newblock {\em IEEE Trans. Audio, Speech, Lang. Process.\/}  {16} (2008).
\newblock


\bibitem[\protect\citeauthoryear{Tzanetakis and Cook}{Tzanetakis and
  Cook}{2002}]%
        {Tzanetakis2002}
{G. Tzanetakis} {and} {P. Cook}. 2002.
\newblock \showarticletitle{Musical genre classification of audio signals}.
\newblock {\em IEEE Trans. Speech Audio Process.\/} {10}, 5 (July 2002),
  293--302.
\newblock


\bibitem[\protect\citeauthoryear{Urbano, Schedl, and Serra}{Urbano
  et~al\mbox{.}}{2013}]%
        {Urbano2013}
{J. Urbano}, {M. Schedl}, {and} {X. Serra}. 2013.
\newblock \showarticletitle{Evaluation in music information retrieval}.
\newblock {\em J. Intell. Info. Systems\/} {41}, 3 (Dec. 2013), 345--369.
\newblock


\bibitem[\protect\citeauthoryear{Wang}{Wang}{2003}]%
        {Wang2003}
{A. Wang}. 2003.
\newblock \showarticletitle{An industrial strength audio search algorithm}. In
  {\em Proc. Int. Soc. Music Info. Retrieval}.
\newblock


\bibitem[\protect\citeauthoryear{Wiggins}{Wiggins}{2009}]%
        {Wiggins2009}
{G.~A. Wiggins}. 2009.
\newblock \showarticletitle{Semantic gap?? {S}chemantic Schmap!!
  {M}ethodological Considerations in the Scientific Study of Music}. In {\em
  Proc. IEEE Int. Symp. Mulitmedia}. 477--482.
\newblock


\bibitem[\protect\citeauthoryear{World Sport Dance Federation}{World Sport
  Dance Federation}{2014}]%
        {WSDF2014}
World Sport Dance Federation 2014.
\newblock {\em World Sport Dance Federation Competition Rules}.
\newblock World Sport Dance Federation, Bucharest, Romania.
\newblock
\showURL{%
\url{https://www.worlddancesport.org/Rule/Competition/General}}


\end{thebibliography}


\received{and accepted Apr. 2015}{Oct. 2015}{and published ????}

%
%
%
%
%
%

\end{document}